\documentclass[natbib,smallcondensed]{svjour3}     
\usepackage{graphicx}
\usepackage{booktabs}  
\usepackage{xspace}
\usepackage{lscape}
\usepackage{bm}
\usepackage{soul}

\usepackage{xcolor}                       
\usepackage[colorlinks=true,citecolor=blue,urlcolor=blue,breaklinks]{hyperref}

\graphicspath{{figs/},{upload1/},{../}}

\journalname{Living Reviews in Solar Physics}

\newcommand{\del}{\mbox{\boldmath $\nabla$}}
\def\vec{\boldsymbol}
\def\avg{\bar}

\graphicspath{{./fig/}{./png/}}
\usepackage{color}
\def\blue{\bf\textcolor{blue}}

\def\bl{Babcock--Leighton}
\def\mc{meridional circulation}
\def\mf{meridional flow}
\def\dr{differential rotation}
\def\mm{Maunder minimum}
\def\gm{grand minima}
\def\go{Gnevyshev--Ohl}
\def\to{torsional oscillation}
\newcommand{\Fig}[1]{Figure~\ref{#1}}
\newcommand{\Figs}[2]{Figures~\ref{#1} and \ref{#2}}

\newcommand{\Eq}[1]{Equation~(\ref{#1})}
\newcommand{\Eqs}[2]{equations~(\ref{#1}) and~(\ref{#2})}

\newcommand{\Sec}[1]{Section~\ref{#1}}
\newcommand{\Secs}[2]{Sections~\ref{#1} and \ref{#2}}

\def\avg{\overline}

\newcommand{\mps}{m~s$^{-1}$}
\newcommand{\cmss}{cm$^2$~s$^{-1}$}

\begin{document}

%
%
%
%
%
%
%

\title{Models for the long-term variations of solar activity%
}

\author{Bidya Binay Karak}

\institute{B. B. Karak \at 
              Department of Physics, Indian Institute of Technology (Banaras Hindu University), Varanasi, India \\
              \email{karak.phy@iitbhu.ac.in}           
}
\date{Received: date / Accepted: date}

\maketitle

\begin{abstract} 
One obvious feature of the solar cycle is its variation from one cycle to another.
In this article, we review the dynamo models for the long-term variations of the solar cycle.
By long-term variations, we mean the cycle modulations beyond the 11-year periodicity and these include,
the Gnevyshev-Ohl/Even-Odd rule, grand minima, grand maxima, Gleissberg cycle, and Suess cycles.
After a brief review of the observed data, we present the dynamo models for the solar cycle.
By carefully analyzing the dynamo models and the observed data, we identify the following broad
causes for the modulation: (i) magnetic feedback on the flow, (ii) stochastic forcing, and (iii) time delays in various processes of the dynamo.
To demonstrate each of these causes, we present the results from some illustrative models for the cycle modulations 
and discuss their strengths and weakness.  We also discuss a few critical issues and their current trends.  
The article ends with a discussion of our current state of ignorance about comparing detailed features of the magnetic cycle
and the large-scale velocity from the dynamo models with robust observations.
\end{abstract}


\setcounter{tocdepth}{3}
\tableofcontents



\section{Introduction}
\label{section:introduction}

The most prominent and fundamental feature of the solar magnetic field is its 11-year cyclic oscillation.
Systematic observations of the large-scale solar magnetic field, 
available since the 1950s, revealed the reversals of the field.
However, the times of the reversals and the strength of the field are not the same for all the cycles.
Time series of sunspot number and the sunspot area, for which we have direct observations for longer durations
(group sunspot number since 1610 and the area since 1874), also show cycle-to-cycle variations \citep{Hat15}; \Fig{fig:obs_ssn}.
Thus, there is no doubt that the 11-year solar cycle is not regular and that makes the prediction of the future cycle
a formidable task \citep{Petrovay20}. 
The prediction, however, is essential as the Sun's magnetic field drives the space weather which sometimes poses serious problems to us (e.g., by damaging satellite's electronics,
modern-day technologies such as telecommunications, GPS networks, and electric power grids at high latitudes,
making polar routes dangerous for aviation, increasing the radiation dose to astronauts in space); \citet{Temmer21}.  
Evidence suggests that the variable solar activity may also drive changes in the Earth's global temperature \citep{Solanki13}.

Simply saying that the solar cycle is irregular is not enough to describe its true nature;
there are many distinct features---such as grand minima and grand maxima---which can be considered as extreme examples
of irregularity; \Fig{fig:ssnC14}. Additionally, there are some long-term patterns beyond the usual 11-year variation,
such as Gnevyshev--Ohl Rule and Gleissberg cycle.
Below we briefly discuss these long-term variations. 
However, the readers can check the excellent reviews \citep{Hat10, Hat15, Uso17, Uso23, Biswas23} for extensive discussion.

\section{Long-term variations of the solar cycle}
\subsection{Grand minima and maxima}
\begin{figure}
\centering
\includegraphics[width=1.0\columnwidth]{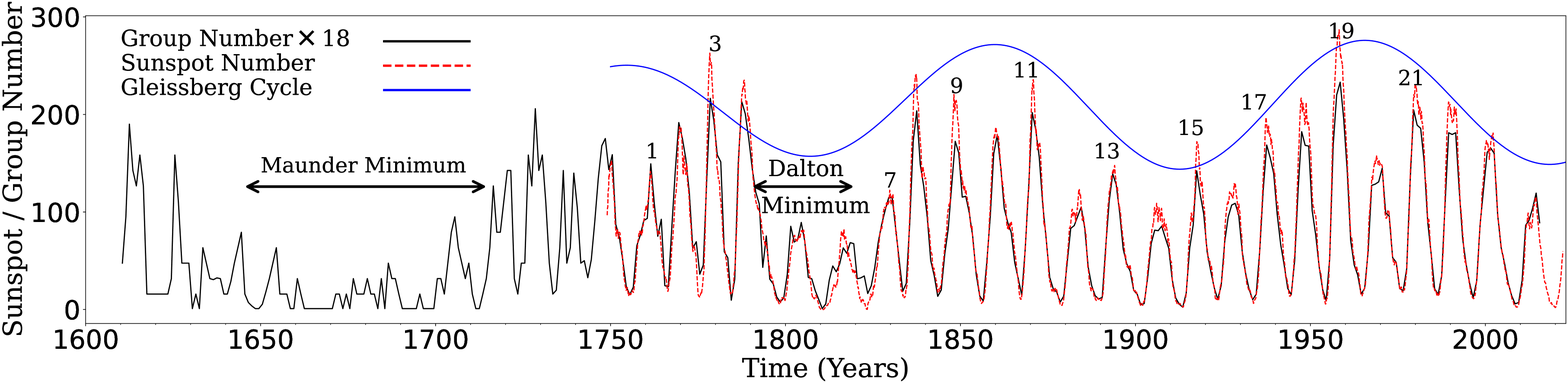}
\caption{Yearly variation of the monthly mean sunspot number smoothed using a Gaussian filter of FWHM $=7$~months (red curve), available since 1749 and the yearly mean group sunspot number (black curve) available during 1610--2015. Note that group number is scaled by a factor of 18 to bring it to
the scale of sunspot number. The blue curve with 98-year period guides the Gleissberg cycle. 
Cycle numbers for which the Even-Odd effect is obeyed are shown by tagging the number on the odd cycles. 
Data source: WDC-SILSO, Royal Observatory of Belgium, Brussels.}
\label{fig:obs_ssn}
\end{figure}

\begin{figure}
\centering
\includegraphics[width=1.05\columnwidth]{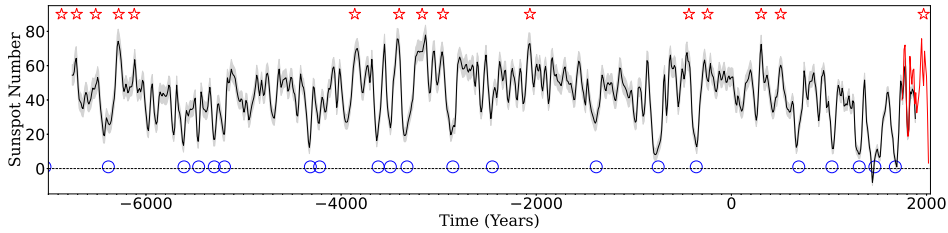}
\caption{Reconstructed (decadal) sunspot number along with its $68\%$ confidence interval (gray shading) over nine millennia derived from $^{14}$C data using a multi-proxy Bayesian method by \citet{Wu18}. The red curve
shows the decadally resampled international sunspot number for the last 300 years (version 2, scaled by 0.6).
 The dashed line marks the zero spot number.
 Asterisks and circles mark the times of occurrences of grand maxima and minima, respectively.
 }
\label{fig:ssnC14}
\end{figure}

Grand minima are the extended episodes of   considerably lower magnetic activity than the normal one.
The best example of these is the Maunder minimum in the 17th century  
when solar activity was considerably weaker than the normal one    for about 70 years \citep{Eddy}; \Fig{fig:obs_ssn}.
We emphasize that this is not an artifact 
due to few observations but a real and well-observed event \citep{HS96}.
Observations have shown that the \mm\ was not 
complete lack 
of activity---the Sun was still producing
spots and even cycles but at a lower rate \citep{BTW98, Zolotova_2015, Uso15, Vaq15, ZP16}. 
This makes the \mm\ (and possibly all grand minima) a special state of solar activity.
Further 
distinct aspects of \mm\ are the followings. (i) There was a strong hemispheric asymmetry during the latter half of \mm;
sunspots were observed mostly in the southern hemisphere \citep{RN93}. (ii) Recovery of \mm\ is gradual, however, the onset is somewhat uncertain (\Fig{fig:obs_ssn}). Some previous results suggested that the onset is abrupt \citep{UMK00}, while later studies showed that it is likely to be gradual \citep{Vaq11}.  
(iii) A proxy of solar activity inferred from a cosmogenic isotope $^{14}$C showed 
that the cycle length before the onset of the \mm\ was significantly longer than 
its usual value \citep{Miy21, Usoskin21}. 

Analyses of $^{14}$C for the last 11,400 years revealed the following important results 
\citep{USK07, Uso17, Usoskin21}. (i) The Sun spent about 
17$\%$ of its time in the grand minimum state. 
(ii) Grand minima are of two types: short minima of Maunder type with duration 30–90 years
and long ($>100$ years) minima of Sp\"orer type. (iii) 
The grand minima recur aperiodically.
(iv) The waiting time distribution of the occurrence of grand minima displays 
a deviation from an exponential distribution. However, \citet{Mea08}
showed that feature (iv) can be an artifact of poor statistics (it is based on only 27 grand minima identified in the observed data).

Grand maxima, on the other hand, are extended periods with appreciably higher magnetic activity than the 
normal one. 
The modern maximum that occurred around 1960 is an example of the same. 
In about last 11,000 years,
23 grand maxima were detected and the Sun spent about $12\%$ of its time in this phase \citep{USK07, Uso17}.  Grand maxima are 
more short-lived than the grand minima  \citep{sol04, Usoskin21}.
The distribution of the duration of grand maxima shows a smooth variation with an 
exponential fall at longer durations.
The distribution of the waiting time between the consecutive grand maxima is 
not conclusive but there is an indication of deviation from the exponential law.

\subsection{Cycles and modulations beyond 11 years}
Beyond the regular 11-year solar cycle, the following longer 
cycles or modulations are detected in the solar activity data.
\begin{itemize}
 \item Gnevyshev--Ohl Rule/Even-Odd effect: This says that if the 
cycles are arranged in pairs with the even cycle and the following odd cycle, 
then the sum of the sunspot number in the odd cycle is higher than the even 
cycle \citep{GO48}. 
We note that this is not a strict rule, it is violated in cycle pairs: 4--5, and 22--23 \citep{Hat15}.
\item 
Gleissberg cycle: a modulation in the solar activity with a mean 
period of about 90 years \citep{glessberg, Hat15}. Recent data shows that a 
sinusoidal fit to the detrended amplitude gives an approximate period of 98 years as shown in \Fig{fig:obs_ssn}; also see \citet{forg04} who found somewhat closer value (104 years) in the area-weighted sunspot group data.
\item 
Suess/de Vries cycle: Cycle with a period of 205--210 years detected in the cosmogenic isotopes \citep{Suess}. 
\end{itemize}
Some other cycles like the 
millennial 
Eddy cycle and 2400-year Hallstatt cycle 
are also noticed in the cosmogenic isotope, however, 
their signals are poor and require longer data to confirm \citep{Uso17}.

We would like to mention that not only the solar cycle is irregular, but the magnetic cycles
that are observed in other sun-like stars are also irregular \citep{BoroSaikia18, garg19}.
Analyzing the data of 111 stars of spectral type F2--M2, \citet{Baliu95} showed that the slowly rotating (old) stars show a smoother variability in the magnetic cycle and possibly occasional grand minima, 
whereas the rapidly rotating young stars show 
irregular activity 
and no grand minima.
Recently, \citet{Shah18} and \citet{Baum22} claim that HD 4915 and HD 166620 are possibly entering into the grand minimum phase. Hence, stellar cycles are also interesting in terms of their modulations. 

Now we shall come to the models for these long-term variations of solar activity. By models in this article, we mean the dynamo models that are used to explain the long-term variations of solar activity. Below we first present an introductory discussion of the solar dynamo (\Sec{sec:dynamo_intro}) and models (\Sec{sec:dynamohistory}). Then we identify the causes of the long-term modulations (\Sec{sec:causes}), followed by some illustrative models for the same (\Secs{sec:mod_variability}{sec:MHDmodels}). 
Finally, we discuss a few open questions with current trends (\Sec{sec:questions}) and end the article with concluding remarks (\Sec{sec:conclusion}).  
 
\section{Solar dynamo: An overview}
\label{sec:dynamo_intro}
Dynamo is a process in which a 
sufficiently strong and complex plasma flow 
maintains a
magnetic field by overcoming its Ohmic dissipation. 
The magnetic fields that give rise to
sunspots, global dipole magnetic field, and 11-year cycle are essentially of the large-scale (global) type which usually requires a non-zero net helicity (mirror asymmetry) in the flow.
Due to the global rotation, the convective motion of plasma in the Sun is helical and the rotation is differential (nonuniform).
Through this helical flow, the poloidal field in the Sun is primarily generated from the toroidal field, 
while the poloidal field 
acts as a source for 
the toroidal field through the differential rotation \citep{Pa55}. 
Thus, the solar dynamo is essentially a cyclic oscillation between two fields and the turbulent transport 
plays an essential role in this oscillation (\Sec{sec:dynamohistory}). 
For a detailed discussion of
the solar dynamo, we refer the readers to the comprehensive reviews by 
\citet{Oss03} and \citet{Cha05,Cha10,Cha20}.

To study the solar dynamo, we need to begin with at least two basic equations of 
the magnetohydrodynamics, namely, 
\begin{equation}
 \frac{\partial \vec{B}}{\partial t} =  \vec{\nabla} \times (\vec{v} \times \vec{B} - \eta \vec{\nabla} \times \vec{B}),
 \label{eq:ind}
\end{equation}
\begin{equation}  
  \label{eq:mom}
  \rho \left[\frac{\partial \vec{v}}{\partial t} + (\vec{v} \cdot \del) \vec{v} \right] =  -\del P + \vec{J}\times\vec{B}+ \del \cdot (2 \nu \rho \vec{S}) + \vec{F},
\end{equation}
where $\vec{B}$ and $\vec{v}$ are the magnetic and velocity fields, respectively, $\eta$ is the magnetic diffusivity, $\rho$ is the density, $P$ is the pressure, $\vec{J} = \del \times \vec{B}/\mu_0$, the current density, $\nu$ is the kinematic viscosity, $S_{ij} = \frac{1}{2} ( \nabla_i v_j + \nabla_j v_i) - \frac{1}{3} \delta_{ij} \del \cdot \vec{v}$ is the 
 rate-of-strain tensor 
and the term $\vec{F}$ includes gravitational, Coriolis and any other body forces acting on the fluid.

The above equations must be solved along with the mass continuity, internal energy and equation of state
in the solar CZ with appropriate boundary conditions. This approach to studying the solar dynamo---so-called global MHD simulations---was pioneered by \citet{Gi83} and \citet{Glatz84} in the 1980s. 
While these simulations gave a few positive results (e.g., large-scale flows and field and a bit of polarity reversal), being computationally expensive, these simulations were hardly applied to explore the long-term variations of the solar cycle. Further, the applicability of these simulations in the Sun is questionable due to their operation in 
a completely different parameter regime. In recent years, however, we have got some encouraging results in 
the global MHD simulations, a few of them were run for several magnetic cycles to explore the cycle modulations.
In \Sec{sec:MHDmodels}, we shall discuss the modulations of cycles found in these simulations. Probably the biggest problem in
these simulations is to explore and understand the cause of solar cycle variabilities. 
In fact, often an equivalent mean-field model is set up to identify dynamics of the magnetic field in these global MHD simulations. 
On the other hand, the mean-field models have been extensively employed in the past to explore the cause
of solar cycle variability. 
Therefore, below we shall consult the mean-field version of the above equations to identify various
mechanisms that can possibly lead to cycle modulations.

Writing the magnetic and velocity fields in terms of mean/large-scale
and fluctuating/small-scale parts and applying suitable approximations in \Eqs{eq:ind}{eq:mom},
one can obtain the following equations \citep{KR80}.
\begin{equation}
 \frac{\partial \vec{\avg B}}{\partial t} =  \vec{\nabla} \times ( \vec{\overline{v}} \times \vec{ \avg B} + \vec{ \overline{\cal E}} ),
 \label{eq:indm}
\end{equation}
	
\begin{equation}  
  \label{eq:momm}
  \rho \left[\frac{\partial \vec{\avg v}}{\partial t} + (\vec{\avg v} \cdot \del) \vec{\avg v} \right] =  -\del P + \vec{\avg J}\times\vec{\avg B} +  \avg{ \vec{ J^\prime} \times \vec{B^\prime} }  - \vec{ \nabla} \cdot \rho \vec{\avg Q} + \vec{\avg F},  
  \label{eq:nsm}
\end{equation}
where the quantities with overline and prime respectively denote the mean and fluctuating components.
The  mean electromotive force  $\vec{ \overline{\cal E}}$  is given by
\begin{equation}
\vec{ \overline{\cal E} }  = \avg{ \vec{   v^\prime  }  \times \vec{ B^\prime } } .
\label{eq:emf_full}
\end{equation}
 In the mean-field theory, this  $\vec{ \overline{\cal E}}$ is written in terms of the mean magnetic field in some limiting case 
 (which holds at small Strouhal and magnetic Reynolds numbers) as follows.
\begin{equation}
       \overline{\cal E}_i  = \hat{ \alpha}_{ij}  \avg B_j + \hat{ \eta}_{ijk} \frac{\partial \avg B_j}{\partial x_k},  
\label{eq:emf_i}
\end{equation}
Some components of $\hat{\alpha}$ tensor give the dynamo action and some of the 
components of $\hat{ \eta}$ tensor are responsible for the diffusion of fields. 
For  homogeneous and isotropic turbulence, one can show that 
\begin{equation}
   \vec{ \overline{\cal E} }  =  \alpha \vec {\avg B} -  \eta_t ( \vec{\nabla} \times \vec {\avg B}),
\label{eq:emf}
\end{equation}
where  $\alpha = - \frac{1}{3} \tau_{\rm corr} (  \avg{ \vec{   v^\prime  }  \cdot \del \times \vec{v^\prime} }  - (\rho \mu_0)^{-1} \avg{ \vec{   B^\prime  }  \cdot \del \times \vec{B^\prime} } )$ and 
$\eta_t \approx \frac{1}{3}  \tau_{\rm corr}  \avg{ \vec{   {v^\prime} }  \cdot  \vec{{v^\prime}} }$. 
In the classical $\alpha$${\rm \Omega}$ dynamo model, this $\alpha$ coefficient is the one which is responsible for the generation
of the poloidal magnetic field from the toroidal one \citep{Pa55, SKR66}. The turbulent diffusivity $\eta_t$ is several orders of
magnitude larger than the molecular $\eta$ and that is the reason we have dropped out the $\eta$ term in \Eq{eq:indm}.
  
If the turbulence is inhomogeneous, then there will be an additional term $\vec{\gamma} \times  \vec {\avg B}$ in the above $\vec{ \overline{\cal E}}$.
This $\vec{\gamma}$ is the magnetic pumping which is usually ignored in most of the 
kinematic mean-field
dynamo models, but found to be important in the solar dynamo \citep[e.g.,][]{GD08, KN12, KO12, Ca12, KC16, KM17} and has also been detected in global convection simulations \citep{RCGBS11, ABMT15, sim16, War18}.

The value of $Q$ in \Eq{eq:nsm} can be written in terms of Reynolds and Maxwell stresses as 
\begin{equation}
      \avg{Q}_{ij}  = \avg{ v_i^\prime v_j^\prime }  - (\rho \mu_0)^{-1}  \avg{ B_i^\prime B_j ^\prime}  =   \avg{Q}_{ij}^\lambda  - {\cal N}_{ijkl}  \frac{\partial \avg v_k  }{\partial x_l},
\label{eq:lamb}
\end{equation}
where the tensor  $\vec{{\cal N}}$ gives the turbulent viscosity $\nu_t$. While $\vec{{\cal N}}$ is rotation dependent and has a complex form, 
 it has two simple terms in the case of isotropic turbulence. 
Again like $\eta_t$, $\nu_t$ is usually much larger than the molecular viscosity ($\nu$) and thus
the latter is neglected in \Eq{eq:nsm}.

The term $\avg{Q}_{ij}^\lambda$ is called the $\Lambda$ effect which drives angular momentum in the rotating CZ \citep{Kipen63, Rudiger89, KR93} to give rise to the differential rotation. While the second term in \Eq{eq:lamb} 
tends to 
smooth out the nonuniformity in rotation, $\avg{Q}_{ij}^\lambda$  
makes the rotation nonuniform.

\section{Some historical developments of the dynamo models}
\label{sec:dynamohistory}
\subsection{Axisymmetric kinematic dynamo equations}

Despite tremendous contribution to the field, the pioneering simulations of \citet{Gi83} and \citet{Glatz84, Glatz85} were discouraging for the dynamo modellers as they failed to reproduce most of the basic features of the solar magnetic field. 
Due to this failure, the dynamo modellers paid more attention to the mean-field models
to study the solar cycle. Motivated by the observed large-scale magnetic and velocity fields, the coronal structure, and to make calculations simple, historically the mean-field solar dynamo was mostly studied under the axisymmetric approximation.
With this approximation, the large-scale magnetic field can be written as
\begin{equation}
\vec{\overline{B}}(r,\theta,t) = \vec{B_{\rm p}} + \vec{B_{\rm t}} =  \del \times \left[A(r,\theta,t)\vec{ \hat{\phi}} \right] + B (r,\theta,t) \vec{ \hat{\phi}},
\end{equation}
where $\vec{B_{\rm p}} = B_r (r,\theta,t)  \vec{ \hat{r}} + B_\theta (r,\theta,t)  \vec{ \hat{\theta}} $ is the poloidal component of the magnetic field  and 
$ \vec{B_{\rm t}} =  B (r,\theta,t) \vec{ \hat{\phi}}$
is the toroidal component.
Similarly, the velocity can be written as
\begin{equation}
\vec{\overline{v}}(r,\theta) = \vec{v_{\rm m}} (r,\theta) +  v_\phi (r,\theta)  \vec{ \hat{\phi}}  =  v_r (r,\theta)  \vec{ \hat{r}} + v_\theta (r,\theta)  \vec{ \hat{\theta}}  + r \sin\theta {\rm \Omega} (r,\theta) \vec{ \hat{\phi}},
\end{equation}
where $ \vec{v_{\rm m}} =  v_r (r,\theta)  \vec{ \hat{r}} + v_\theta (r,\theta)  \vec{ \hat{\theta}} $ is the meridional circulation and ${\rm \Omega} (r,\theta)$ is the angular frequency.   

We note that here  
we have taken the velocity as the time-independent (steady), which is the case when we make the kinematic approximation. The kinematic approach has been adopted extensively in the literature to study the solar dynamo
because in this case, we do not have to consider the equation for the flow (\Eq{eq:nsm}) and thus it makes the dynamo problem linear (see \Eqs{eq:pol}{eq:tor} below) and thus simpler.  Nevertheless, observations provide us with the azimuthal flow in the whole CZ and the meridional flow in the near-surface layer. Given the fact that the differential rotation shows a little variation (in the form of torsional oscillation), one would expect that the kinematic approach is not a bad assumption for the sun.

After substituting the above forms of the fields in \Eq{eq:indm} and using the value of $\overline{\cal E}$ from \Eq{eq:emf} one can derive the following equations.
\begin{equation}
\frac{\partial A}{\partial t} + \frac{1}{s}(\vec {v_m} \cdot \del)(s A)   = \eta_t\left(\nabla^2 - \frac{1}{s^2}\right)A + \alpha B,
\label{eq:pol}
\end{equation} 
\begin{equation}
\frac{\partial B}{\partial t} + \frac{1}{r}\left[\frac{\partial (r v_r B)}{\partial r}+ \frac{\partial (v_\theta B)}{\partial \theta}  \right] = \eta_t\left(\nabla^2 - \frac{1}{s^2}\right)B 
+ s(\vec {B_p} \cdot \del){\rm \Omega} + \frac{1}{r}\frac{d\eta _t}{dr}\frac{\partial (rB)}{\partial r},
\label{eq:tor}
\end{equation}
where $s= r\sin{\theta}$ and $\eta_t$ is assumed to depend only on $r$.

In the above equations, the second term involving $\vec{v_m}$ (or equivalently $v_r, v_\theta$) corresponds to advection of poloidal field by meridional flow and the first terms on the RHS of both equations represent the diffusion. 
 In \Eq{eq:pol}, the term $\alpha B$ is the source for the poloidal magnetic field. We discuss more about it 
in the next section.  
The term $s(\vec{ B_p} \cdot \del){\rm \Omega}$ in \Eq{eq:tor} is the source for the toroidal field, in which the nonuniformity of the rotation along the direction of poloidal field induces a toroidal field (the ${\rm \Omega}$ effect). 
We note that while deriving above equations we have 
neglected a term $\vec{\hat{\phi}} \cdot [\del \times (\alpha \vec{B_p})]$. This is also a source for the toroidal field through the $\alpha$ effect. However,
in Sun we believe that this term is negligible compared to the source due to differential rotation.
The dynamo model constructed based on this assumption is traditionally called the $\alpha$${\rm \Omega}$ type dynamo in which the poloidal and toroidal fields maintain each other through a feedback loop.
Finally, the last term in \Eq{eq:tor} gives rise to an advection 
due to nonuniform turbulent diffusion. 

\begin{figure}
\centering
\includegraphics[scale=0.20]{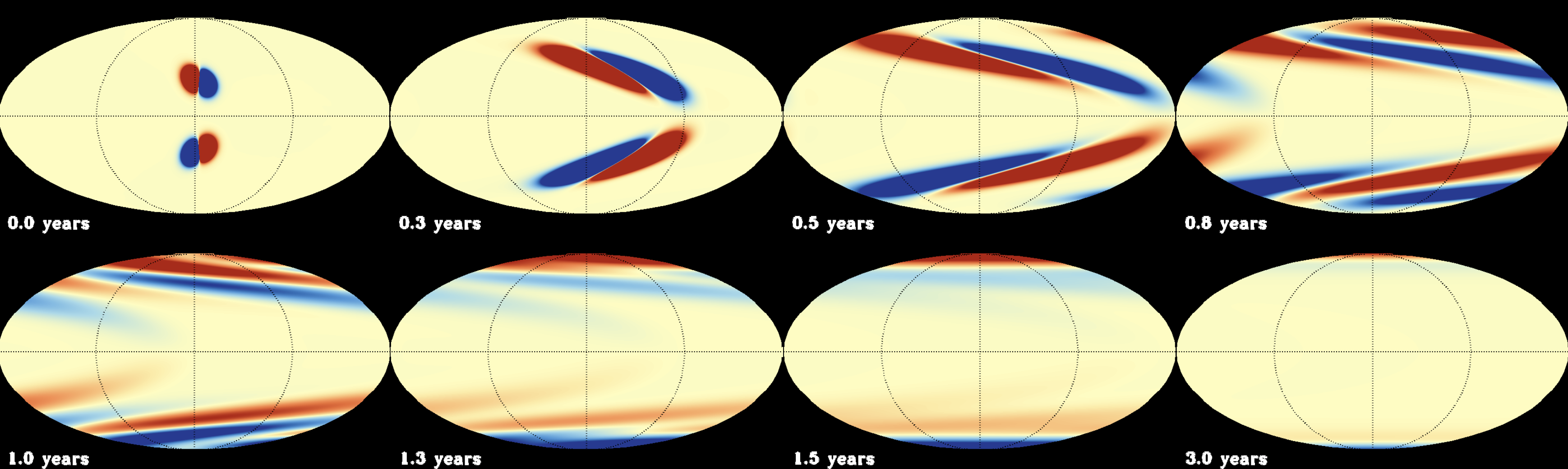}
\caption{Demonstration of \bl\ process. Decay and dispersal of two BMRs deposited symmetrically at $25^\circ$ are shown for three years. Note that due to finite tilts ($\approx 14^\circ$ as assigned by Joy's law) of the BMRs, a net poloidal field near the pole is produced (see the weak field near the pole in the last snapshot). Snapshots are taken from a 3D model \citep{KM18} in which meridional flow (poleward on the surface), differential rotation (fast equator) and a turbulent diffusivity of $10^{12}$~\cmss are specified. While the magnetic field in the initial BMR is 3000~G, it is saturated at 0.1~G to show the weak field at the end.}
\label{fig:BLmechanism}
\end{figure}

\subsection{\bl\ dynamo models}
There are two potential and widely studied mechanisms through which the poloidal field in the Sun can be generated. 
The first one was proposed by \citet{Pa55} and mathematically studied in detail by \citet{SKR66}. In this process, the helical convection (which has non-zero net helicity) in the CZ twists the toroidal field to generate a poloidal component. 
This is the traditional $\alpha$ effect.
We note that this process only operates when the energy density of the toroidal magnetic field is below the energy density of the convective turbulence so that the turbulent flow can twist the field. 
The second mechanism goes to \citet{Ba61} and \citet{Le64} who proposed that the decay of tilted bipolar magnetic regions (BMRs) generates a poloidal field. 

Observations show that the line joining the centres of poles of a BMR makes an angle with respect to the equatorial line. 
This tilt statistically increases with the latitude which is known as Joy's law \citep{Hale19}.
When a BMR decays, the magnetic field of two poles diffuses at slightly different latitudes.
The field from the leading part 
predominantly cancels with 
the field from the other 
hemisphere across the equator and the 
trailing part is largely advected towards the pole to produce a global dipole moment for the Sun; see \Fig{fig:BLmechanism} and also see \citet{HCM17}. This process of generating poloidal field, the so-called \bl\ process requires a tilt 
in the BMR and this process need not be very efficient in terms of the fact that the magnetic flux in the polar cap 
is approximately the net flux content in only one BMR \citep[$\approx 5\times 10^{21}$~Mx;][]{CS15}. 

As the BMRs are produced from the deep-seated toroidal field and the decay of BMRs produces a poloidal field, this mechanism in many \bl\ type dynamo models is phenomenologically 
prescribed by a term $\alpha B$ in \Eq{eq:pol}. Here, $\alpha$ is nonzero only near the surface in contrast to the (classical) $\alpha$ effect, which operates in the whole CZ.
Although both the $\alpha$ effect and \bl\ process for the generation of the poloidal field are captured by the same term $\alpha B$ in mean-field dynamo models, their mechanisms are completely different. Nevertheless, in 
some 
comprehensive dynamo models, e.g., the 3D model of \citet{YM13, MD14, Kumar19, BC22} and the $2\times2$D model of \citet{LC17}, do not capture the \bl\ process through this $\alpha B$ term, instead, they include explicit BMRs which self-consistently generates the poloidal field.
We further emphasise that the  \bl\ dynamos are also of mean-field type although the \bl\ source term is often introduced in a heuristic way instead of strictly deriving it from the mean-field MHD.

There was little doubt, at least, after the helioseismology mapped the rotation profile in the CZ \citep{How09} that the 
toroidal field in the Sun is produced through the ${\rm \Omega}$ effect. However, it was not clear 
whether the helical $\alpha$ effect or \bl\ process is the dominating mechanism for the generation of
the poloidal field in the Sun. Most of the early dynamo models considered the $\alpha$ effect. 
Then in the 1990s, 
 the studies from the thin flux tube model have shown that the tilt of the BMR is presumably produced due to the Coriolis force acting on the rising flux tube of the toroidal field in the CZ \citep{DC93, FFM94}. 
 These studies also showed that the magnetic field at the bottom of the CZ (BCZ) in the BMR forming regions\footnote{During the BMR formation, the magnetic field may also be locally amplified from the diffuse field; see discussion in \citet{Getling19}.}
  has to be of the order of $10^5$~G, which is much higher than the equipartition field strength \citep[$\sim 5\times 10^3 $~G;][]{Stix02}. 
If this is the case, then the 
classical $\alpha$ will not be able to operate in this strong field. However, the \bl\ process still can operate in this regime 
as in this process, the original toroidal field and the generated poloidal field are segregated in space and there is no generation of magnetic helicity and no catastrophic quenching \citep{KO11c}.
This was a strong support behind the \bl\ process as the primary source for the poloidal field in the Sun. 

We would like to recall that there are observational supports for the operation of the \bl\ process. First, there is a close connection between the locations of BMR emergences and the sites of the formations of the trailing polarity surges at low latitudes, as seen in the time-latitude distribution (or maybe in the Carrington maps) of the magnetic field \citep[e.g.,][]{Mord20,Mord22}.  Second, the Surface Flux Transport (SFT) models, which describe the evolution of the radial component of the magnetic field by utilizing the observed
BMRs, large-scale flows (such as differential rotation and meridional
circulation) and turbulent diffusion on the solar surface, capture the \bl\ process 
\citep{Wang1989, Bau04, UH14, Ji14}. The remarkable success of the SFT model 
in terms of reproducing the observed magnetic field on the solar surface 
as well as the coronal structure gives support that at least the observed magnetic field 
in the Sun is largely due to the \bl\ process. Next, the observed correlation
between the polar field (or its proxy) at the end of the cycle and the amplitude of the next cycle \citep{WS09, KO11, Muno13, Priy14, Kumar21, Kumar22} and the flux budgets of the observed and the generated poloidal and toroidal fields \citep{CS15} suggest that the \bl\ process is possibly the main source of the poloidal field in the Sun. 

\subsection{Flux transport dynamo models}
While developing a satisfactory dynamo model for the solar cycle in the \bl\ framework, 
modellers faced a challenge to 
reproduce the observed migration of sunspots. As the deep-seated toroidal field gives rise to sunspots and the latitudinal band of sunspots migrates towards the equator within a cycle, we expect the toroidal field in the deeper CZ to advect towards the equator with the progress of a cycle. However, to obtain an equatorward propagation through dynamo wave,  
 Parker-Yoshimura sign rule demands that $\alpha \frac{\partial {\rm \Omega} }{\partial r} < 0 $ 
 in the northern hemisphere \citep{Pa55, Yo75}. The observed profile of ${\rm \Omega}$ shows that  
 $\frac{\partial {\rm \Omega} }{\partial r}$ is positive in the low latitudes where sunspots emerge. Furthermore, the observations of BMR tilts 
 also suggest that the $\alpha$ corresponding to the \bl\ process is also positive in the northern hemisphere. Hence,
 Parker-Yoshimura sign rule suggests a poleward migration in contrast to observations.
 This problem was resolved by introducing a meridional flow such that it is equatorward 
 in the deeper CZ. A sufficiently strong flow can overpower the poleward dynamo wave 
 and can explain the equatorward migration of the sunspot belt \citep{WSN91, CSD95, Dur95, HKC14}. 
 While the poleward component of 
 the meridional flow on the surface 
 has been well-known for many years, recent helioseismic observations
 find 
some 
indication of the equatorward (return) flow near the base of CZ \citep{RA15, Gizon20}. 
Also, there is controversy about the depth of the return         flow and the number of cells that exist in the CZ.

The dynamo models in which the equatorward migration of the toroidal field at the BCZ is driven by 
the meridional flow (or some other flow), rather than dynamo waves,
are popularly known as the flux transport dynamo models \citep{WSN91,CSD95,Dur95}; see \citet{Kar14a} for a review on this topic.  Usually, the \bl\ dynamo models (in which the poloidal source is due to the \bl\ process) include a meridional flow to produce the equatorward migration of the toroidal field and thus they are of flux transport type, however, an $\alpha$${\rm\Omega}$ type dynamo can also be of flux transport type if a sufficiently strong meridional flow is present.  Usually, these flux transport dynamo models, consider a single cell (in each hemisphere) \mc\ profile 
with a return flow of about a few meters per second at a depth of about $0.7R_\odot$.
Most of the existing Babcock–Leighton type flux transport dynamo models are kinematic, however, see \citet{Remp06, Fadil17, BC22} for exceptions.  

\section{Mechanisms of long-term variations}
\label{sec:causes}
With the above introduction to the solar dynamo theory, we shall explore how the irregular variations in the solar cycle can occur. Broadly, we can think of the following three major causes  for these.
\begin{itemize}
 \item Magnetic feedback on the flow
\item Stochastic forcing 
\item Time delays in various processes of the dynamo
 \end{itemize}

\subsection{Magnetic feedback on the flow}
\label{sec:causes_nonlinear}
We have seen in \Sec{sec:dynamo_intro} that the flows are essential for the dynamo mechanism.  The large-scale component of velocity, as appearing in \Eq{eq:indm}, is observed in the form of differential rotation and \mc\ in the Sun. The differential rotation 
induces 
a strong toroidal field from the poloidal one in the CZ. The meridional circulation transports the magnetic field near the surface from low to high latitudes where it pushes the field to the deeper CZ, and it possibly transports the field towards the equator near the BCZ.
Hence, it is natural that any dynamical change in these large-scale flows can cause variation in the solar cycle.

In \Eq{eq:nsm}, we find several dynamical terms through which modulation in the flow can arise. First is the mean Lorentz force $\vec{\overline {J}} \times \vec{ \overline {B}}$, which arises through the interaction between the mean magnetic and the mean current. 
This is also called the Malkus-Proctor effect \citep{MP75}. The second is the small-scale feedback, which consists of two parts. One is the direct small-scale Lorentz forcing 
$\overline {\vec{J^\prime} \times \vec{B^\prime}}$ appearing in \Eq{eq:nsm} (through the fluctuating current and magnetic field). Another is the dynamical modulation in the $\Lambda$ effect, which comes from the anisotropic turbulence (appearing through $\overline{Q}$ in \Eq{eq:lamb}). This modulation arises because the mean magnetic field also gives rise to the Lorentz force on the small-scale turbulence \citep{Kit94_lambda}. This, so-called micro (small-scale) feedback, has been captured through a simple quenching in the $\Lambda$~effect in many mean-field dynamo models \citep{kuker99}. 
 The dynamo-induced small-scale magnetic field also affects the large-scale flows and the turbulent transport; see Equations (\ref{eq:nsm}, \ref{eq:emf}, \ref{eq:lamb}) and \citet{Kap19}.
  
Next, the magnetic field gives feedback on the dynamo coefficients. The magnetic field dependence of the $\alpha$ coefficient is popularly studied in the literature \citep{Pouquet76, FB02, SB04}. While we do not have an analytical theory for the magnetic field dependence of the dynamo coefficients from the first principle, \citet{RK93} and  \citet{KPR94} respectively gave the dependences of the $\alpha$ and $\eta$ on the magnetic field using the quasi-linear and quasi-isotropic turbulence. Based on this theory, when the magnetic field is much larger than the equipartition field strength, the $\alpha$ falls 
as $1/B^3$, while many kinematic dynamo models traditionally use a quenching factor $1 / \left( 1 + ( B / B_{\rm eq})^2\right)$ with $B_{\rm eq}$ being the equipartition field. MHD turbulent \citep{KB09, Kar14b} and global convection simulations \citep{RCGBS11, sim16, War18} also do show some magnetic quenching in $\alpha$. 
Whatever be the exact magnetic field-dependent form of $\alpha$, in all these results one thing is clear: 
as the magnetic field tries to grow, it suppresses $\alpha$ 
and this in turn reduces the generation of the magnetic field. 
Hence, the nonlinearity in $\alpha$ tries to make the cycle regular, rather than 
producing irregularity in the cycle, provided the strength of $\alpha$ is not much higher than the critical $\alpha$.  
   
The feedback on the small-scale turbulence can also be seen in the modulation of turbulent viscosity and magnetic diffusivity, which can in turn give 
some
variation in the magnetic field. However, due to difficulties in computing these turbulent coefficients, we have limited knowledge on how much cycle modulation can arise due to magnetic feedback on the turbulent coefficients; anyhow, see the results of quasi-linear approximation \citep{KPR94} and the convection simulations cited above. Unlike $\alpha$ quenching, the diffusivity quenching, however, 
tends to 
make the model unstable by increasing the magnetic field when the field strength is large \citep{KO10} and thus unless some other mechanism, say $\alpha$ quenching is included, the dynamo usually does not produce stable cycle \citep{KO10, Vindya21}. 
The last two references have also shown that the nonlinearities in $\alpha$ and $\eta$ produce dynamo hysteresis---strong oscillatory magnetic field in the subcritical regime if the dynamo is started with a strong field and decaying solution otherwise if started with a weak field. 
This was also confirmed in numerical simulations of turbulent dynamos \citep{KKB15, Oliv21}.    

We would like to emphasize that the recent observations of stellar rotation \citep{Met16} show that the rate of solar rotation is close to the minimum rate for the onset of the large-scale dynamo \citep[also see][]{R84}. Furthermore, \citet{CS17, CS19} showed that the variability seen 
 in the cosmogenic isotope for the last 10,000~years 
is consistent with the results from the generic normal form model for a noisy and weakly nonlinear limit cycle. Grand minima are only produced when the dynamo is not highly supercritical and the Sun and solar-like slowly rotating stars do produce grand minima \citep{KO10, kumar21b, Vindya21, V23}.
All these suggest that the solar dynamo is only slightly supercritical and weakly nonlinear.  Thus, possibly the nonlinear effects are not very 
important in producing long-term modulations in the solar cycle.

\subsection{Stochastic forcing}
\label{sec:cause_stochastic}
 Solar CZ is highly turbulent and the turbulent quantities (appearing in \Eqs{eq:emf}{eq:lamb}) are subjected to 
fluctuations around their means in a time scale equal to the correlation time of the turbulent convection. As there is a finite number of convection cells over the longitudes at a given latitude in the Sun, the fluctuations in the turbulent coefficients are significant compared to their mean values.   \citet{H88} argued that the fluctuations in the $\alpha$ effect can be larger than its mean and thus they can produce variation in 
the solar cycle \citep{C92, H93, OHS96}.  In fact, due to small-scale dynamo, there are always fluctuations around $\overline{\cal E}$ \citep{Br08, BS08}.
The fluctuations in the angular momentum transport (as parameterized by the $\Lambda$~effect; \Eq{eq:lamb}) can also alter the differential rotation 
and \mc\ and thus can produce modulations in the solar cycle \citep{Remple05b, Fadil17}. 
 

 \begin{figure}
\centering
\includegraphics[scale=0.50]{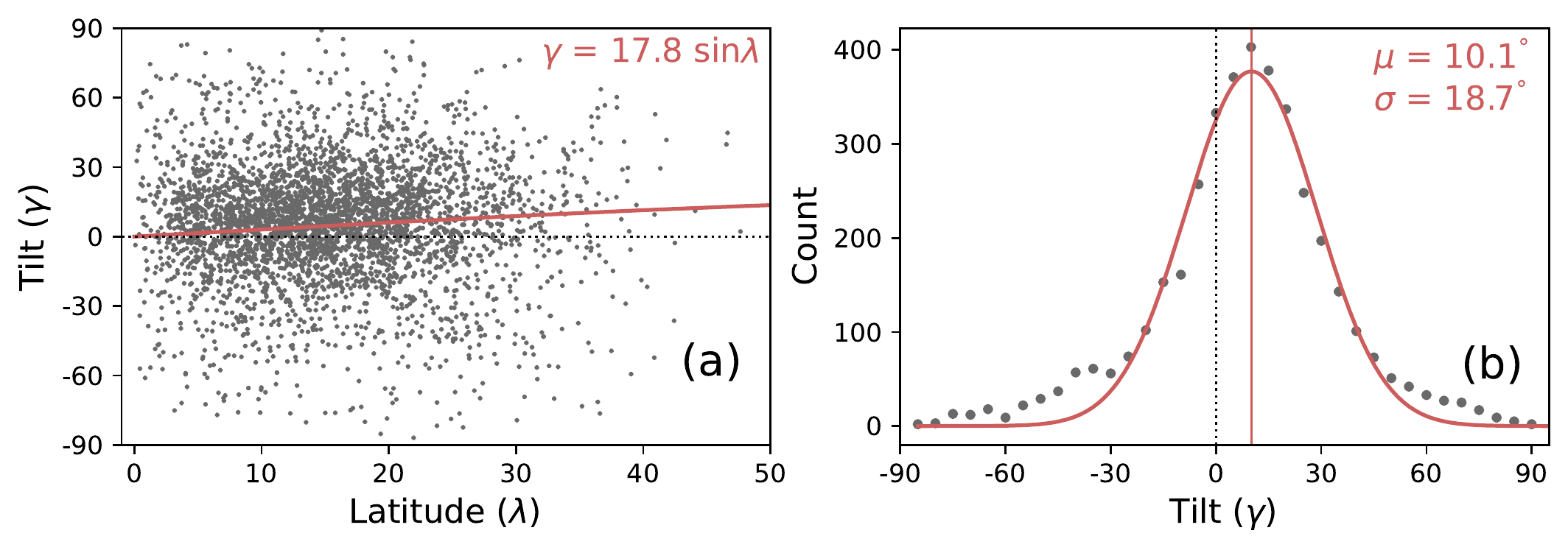}
\caption{Tilt angles of BMRs computed by ``tracking'' the MDI line-of-sight magnetograms covering the Cycle~23 (1996 September--2008 December). Here each BMRs are tracked over their lifetimes 
and the tilt and latitude of a BMR are taken by averaging their values 
over its time evolution when the flux is more than $60\%$ of its maximum. 
In (a) solid line guides Joy's law: $\gamma = 17.8 \sin \lambda$.
(b) Shows the tilt distribution (with $5^\circ$ bin size) with fitted Gaussian (solid line) 
of $\mu = 10.1^\circ$ and $\sigma = 18.7^\circ$. 
The figure is produced using the data presented in \citet{Anu23}.
}
\label{fig:tiltscatter}
\end{figure}

\bl\ process, in which decay and dispersal of tilted BMRs generate a poloidal field in the Sun, involves some intrinsic fluctuations. The tilts of BMRs have a considerable amount of scatter around Joy's law \citep{How91, SK12, MNL14, pavai15, Jha20}. As seen in \Fig{fig:tiltscatter}, the scatter is indeed much larger than the mean. Also, a large number of BMRs are having opposite tilts (negative in the northern hemisphere) due to non-Joy and anti-Hale configurations which generate opposite polarity field.  
Not only the tilt, but the rate of emergence and flux content of BMR also have considerable variations around their means. The cumulative effect of the fluctuations of all these parameters of BMRs can have a large impact on the polar field or the dipole moment at the end of a cycle which can lead to a considerable variation in the solar cycle \citep{Nagy17}.  On average, in the Sun, only a few (new) BMRs per day are produced and thus the short-term variation in the poloidal field is considerably large. This we can also identify by carefully observing the magnetic field on the solar surface \citep{Ca13, JCS14, Mord16, Kit18, KMB18, Mord22}.  The variation in the inflows around BMR \citep{Jiang10, MC17, Nagy20} and the meridional flow  \citep{Bau04, Kar10, UH14b} also can change the amount of poloidal field generated. Like fluctuations in the classical-$\alpha$, there is a long list of work which  reported the fluctuations in the \bl\ process and have utilized these in the dynamo models to reproduce various aspects of the long-term modulation of the solar cycle \citep[e.g.,][]{CD00, Cha04, Ch05, Cha07, CK09, KC11, CK12, OK13, Pas14, LC17, KM17, Nagy17}.

\subsection{Time delay in various processes of the dynamo}
\label{sec:timedelay}
 Time delays are involved in various processes of the dynamo action. 
 \citet{Yo78} argued that the adjustment of the velocity field due to the back reaction of the dynamo-generated magnetic field is not instantaneous and involves a bit of delay---at low Prandtl number, the fluctuations in the large-scale flow lag behind the Lorentz force.
Furthermore, the modification of the thermodynamics due to magnetic feedback alters the velocity field and this involves again a time delay.    \citet{Yo78} showed that a time delay in the dynamo model produces a long-term modulation with occasional ceased activity.  In the \bl\ dynamo framework, some time delays are unavoidable because the sources for the poloidal
 and toroidal fields are spatially segregated.  The poloidal field from the surface layer has to be transported down to the deeper CZ to be sheared 
 by the differential rotation, thus there involves a long time lag between the poloidal to the toroidal field. This lag is comparable to the solar cycle. 
Similarly the toroidal to the poloidal field conversion process also involves a time lag as the toroidal field needs to rise to the surface to form BMRs and then BMRs decay to give rise to the poloidal field. This delay however is short compared to the solar cycle length as the BMR eruption takes a few days to month and the decay takes another few months. 
While these two lags are captured by default in the numerical dynamo models of  \bl\ and interface \citep[in cases where the regions of shear and $\alpha$ effect are spatially segregated][]{MC97} types, 
  a time delay is included by hand in the iterative map and the time delay dynamo models \citep[Sec. 5.4 of][]{Cha10}. 
  This time delay in the {\blue{\textbf {\it nonlinear} }} dynamo model produces a variety of cycle modulations, including 
  Gnevyshev--Ohl rule and intermittent cycles like the grand minima 
 in certain parameter regimes \citep{Dur00, Cha01, wilsmith, Cha07}. 
 While in most of the \bl\ models, the toroidal to poloidal process is assumed to be instantaneous, 
 \citet{Jouve10, Fournier18} included a short delay in their model and made it magnetic field dependent regarding the fact that  
 the flux tube with a strong magnetic field rises fast due to high magnetic buoyancy. 
This  
 magnetic field-dependent time delay during the flux emergence in their flux transport dynamo 
with nonlinear $\alpha$ effect  can produce some 
modulation in the cycle amplitude.
We here note that time delays in all these models produce cycle variability  
only when the nonlinearity becomes important;  
 it is the nonlinearity which is essential to produce cycle modulation. Thus, the time delay alone cannot produce a variability in the solar cycle.

With these basic discussions of the causes of the modulation of the solar cycle, we are now ready to discuss some illustrative models for the long-term variation in the solar cycle.
 
\section{Mean-field models for long-term cycle variabilities}
\label{sec:mod_variability}

\subsection{Models with nonlinear feedback on the large-scale flows}
As discussed in \Sec{sec:causes_nonlinear}, the large-scale flows are subject to change dynamically due to direct Lorentz feedback on the flow \citep{MP75} or through the feedback on the angular momentum transport (like $\Lambda$ effect; \citet{Kit94_lambda}). 
Extensive research has been done on this topic to capture the Lorentz feedback of the dynamo-generated magnetic field 
\citep{Spi77, Tav78, Ruz81}. 
Particularly, using a simplified dynamo model \citet{kuker99} showed that a 
 modification of the differential rotation by the large-scale Lorentz force produces strong modulation in the magnetic cycle including grand minima \citep[also see][]{MB00}.
However, when the $\alpha$ quenching is added, the modulation is drastically suppressed (which is expected as the $\alpha$ quenching 
tends to 
limit the growth of the magnetic field). Some amount of cycle modulation and grand minima are again recovered if a strong $\Lambda$ quenching is included in this model; also see \citet{1999SoPh..189..227K} for a similar study.  In a somewhat improved dynamo model but by including only the feedback of the large-scale magnetic field on the \dr, \citet{Bushby06} find some modulation in the magnetic cycle including grand minima like phases.  
Chaotic solutions are also produced in the highly truncated dynamo model \citep{WCJ84} which produces modulation in the cycle.

\begin{figure}
 \centering
\includegraphics[scale=0.38]{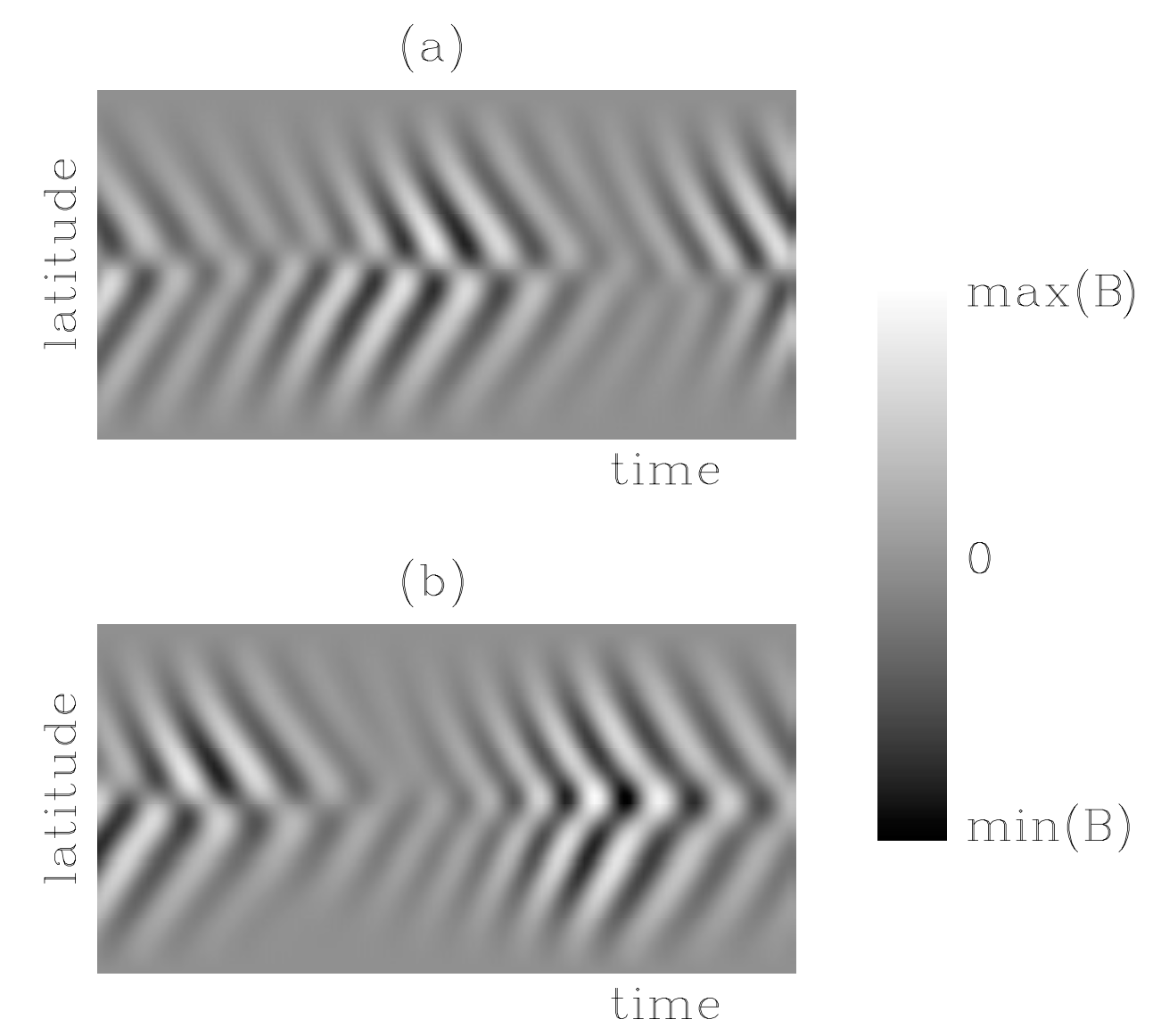}
\caption{Butterfly diagrams from \citet{BTW98}, showing the toroidal field at a fixed radius 
as a function of time and latitude. In (a) the parity is interrupted by the occurrence of a grand minimum. The dynamo recovers from the grand minimum with a strong hemispheric asymmetry.  (b) The grand minimum triggers a  flip from a dipolar to quadrupolar parity. Figure reproduced with permission from author.
}
\label{fig:beer}
\end{figure}

Basically, in all these models, the modulations happen in two ways \citep{KTW98}. In one, the large-scale magnetic field of one parity
(dipolar or quadrupole) drives velocity perturbations and energy is exchanged between the magnetic field and the flow.
In this case, a large variation in the flow velocity is observed with no change in the parity. In the second case,
there exists a nonlinear interaction between the dipole and quadrupole modes, mediated via the velocity perturbation which is driven by the  
Lorentz force. 
This modulation is associated with changes in the parity with almost no change in the velocity \citep{Thelen00}.  These two mechanisms of cycle modulation in the literature are referred as Type II and I, respectively  \citep[e.g.,][]{Tob97,KTW98}.  Based on a nonlinear extension of the \citet{Pa93} model, \citet{BTW98} showed that the latter type of modulation is the cause of Maunder-like grand minima. In \Fig{fig:beer}, we see that during the grand minimum the parity of the magnetic field 
is changed.  Based on a highly idealized simple model of the nonlinear dynamo equations, \citet{WT16} and \citet{BTW18} showed that the long time-scale `supermodulation' apparent in the cosmogenic isotope data can be ascribed to switching of the dynamo between two different modulational patterns i.e., from dipole or quadrupole symmetry to mixed-mode solutions.

Nevertheless, these models are still preliminary and fail to produce many detailed features of the observed magnetic field and the large-scale flows, particularly the correct amount of variation in the differential rotation. Furthermore, in some studies, the amount of feedback on the \dr\ and \mc\ are tuned.

\subsubsection{Variation in \dr}
Observations find almost no variation in the \dr\ \citep{GH84,Jha21} except a tiny one ($< 0.5\%$) around the mean, which is known as the torsional oscillation \citep{How09}. Early dynamo models tried to explain \to\ using the variation of Reynolds stresses due to dynamo-generated magnetic field \citep{kuker96}. Some
other models tried to explain it using the mean Lorentz force of the dynamo-generated magnetic field on the 
momentum equation \citep{Schu81, CCC09}; see \citet{PK19} who included both magnetic feedbacks on the turbulent angular momentum transport and the large-scale Lorentz force.  In spite of that, none of these models could successfully
explain both the equatorward and poleward branches of the \to.
A comprehensive model of \citet{Remp06} showed that an enhanced surface cooling of the active region belt as proposed by \citet{Spruit03}, in addition to the Lorentz forcing, is needed to explain the equatorward branch of \to. This model expectedly finds almost no long-term modulation in the magnetic cycle due to this tiny variation in the \dr. Thus, we may expect that the observed tiny change in the differential rotation may not be a potential cause of the long-term modulation in the solar cycle. However, a series of mean-field dynamo calculations have demonstrated that a variety of cycle modulations including grand minima can be produced due to 
the nonlinear back reaction of the magnetic field on the large-scale flow through the so-called Type~I modulation, which leaves a little imprint in the differential rotation \citep[e.g.,][]{BTW98, KTW98, Bushby06, WT16}. Therefore, it is subtle to answer how much is the role of the tiny variation in differential rotation in producing cycle irregularity.
\subsubsection{Variation in \mc}
The meridional circulation is also subject to vary due to the Lorentz forcing of the dynamo-generated 
magnetic field acting directly on it or through the alteration of the 
differential rotation.
Models including the magnetic feedback often find a cyclic variation in the meridional flow \citep{Remp06, Passos12, HC17}, in agreement with some observations \citep{HR10}.  
Inflows around active regions also cause a cyclic perturbation in the meridional flow \citep{GR08,Gonz08,Gonz10}. 
  This cyclic change in the \mc\ with no overall modulation cannot produce much variation in the magnetic cycle \citep{KC12}. However, when the amount of perturbation in \mf\ varies with the solar cycle strength, it can cause a significant modulation in the solar cycle \citep{Jiang10}. 
Observations find some temporal variations in the \mc\ \citep{Gonz06}, although there is no consensus on its long-term trend due to limited data.  
If the steady-state \mc\ is maintained by a slight imbalance between two large terms---the non-conservative part of the centrifugal force and the baroclinic forcing (which arises due to a latitudinal temperature difference), a slight change in the balance can produce a large variation in the \mc\ \citep{KR99}. 
In fact, the global convection simulations do find a tiny variation in the \dr\ but a considerable variation in the \mc\ \citep{Kar15, Passos17}, somewhat consistent with the available observations.  
By assimilating the synthetic magnetic proxies in the variational data assimilation method based on flux transport dynamo model, \citet{Hung15,Hung17} also find a time-varying \mc.

\begin{figure}
 \centering
\includegraphics[scale=0.40]{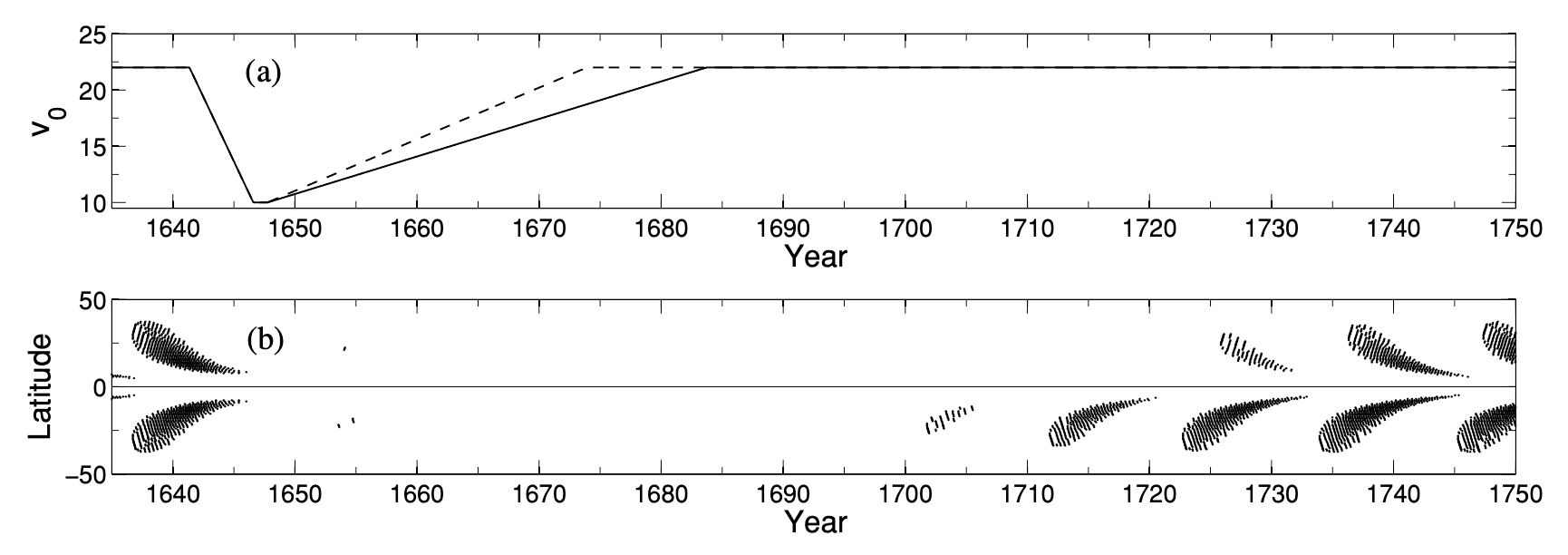}
\caption{Figure showing that a sufficient drop in the \mf\ can trigger a Maunder-like grand minimum. (a) Shows the required \mc\ speed in \mps\ (solid/dashed for north/south). (b) Shows the location of the sunspots from the dynamo model; see \citet{Kar10} for details.}
\label{fig:karak10}
\end{figure}

In the models, particularly in the flux transport dynamo models, the variation in the \mc\ has been found to produce a profound effect on the solar cycle.  In these models, the meridional flow regulates the cycle duration, weaker flow makes the cycle longer and vice versa \citep{DC99}. The \mc\ has also an effect on the cycle strength, however, the effect depends on the diffusivity used in the model. If the dynamo operates in the diffusion-dominated regime (relative importance of the diffusion is more with respect to the advection due to flow), then a weaker flow allows the poloidal magnetic field to diffuse for a longer time and thus makes the magnetic field weak \citep{YNM08}. The opposite scenario happens when the flow is strong. Using this idea, \citet{Kar10}
discovered that if we want to match the cycle duration by adjusting the speed of the meridional flow, then the cycle amplitudes are also matched up to some extent. Thus a significant part of the variation of the solar cycle can easily be modelled simply by varying the speed of the meridional flow. \citet{Kar10} also showed that a sudden weakening of the meridional flow can trigger a
Maunder-like grand minimum as shown in \Fig{fig:karak10}.

Although the results from some of the flux transport dynamo models with variation in the \mf\ 
are very promising, in terms of modelling long-term variations in the solar cycle, it remains to be answered whether there was any large variation in the meridional flow in the past, particularly during the \mm. 

\subsubsection{Joint models with multiple nonlinearities}
A few mean-field dynamo models were developed by considering full MHD equations
with multiple possible nonlinearities \citep{Bra89b, Bra91, BM94, Thelen00, Jennings93, Muhli95,  Remp06, PK19, SM19}. Most of these models included $\alpha$ quenching and Lorentz 
 force feedback (in some form) in the momentum equation.  The aim of these models was
mostly to study the nonlinear stability and the operation of the dynamo.  These models do not produce a considerable long-term modulation and grand minima unless some stochastic fluctuations in the dynamo parameter are included \citep{Fadil17}. This will be discussed in the later sections.


\subsection{Models with fluctuations}
As discussed in \Sec{sec:cause_stochastic}, stochastic fluctuations in the solar dynamo is unavoidable and thus using these fluctuations numerous dynamo models have been constructed to explain the variable solar cycle.

\subsubsection{Fluctuations in $\alpha$ effect}

\begin{figure}
 \centering
\includegraphics[scale=0.40]{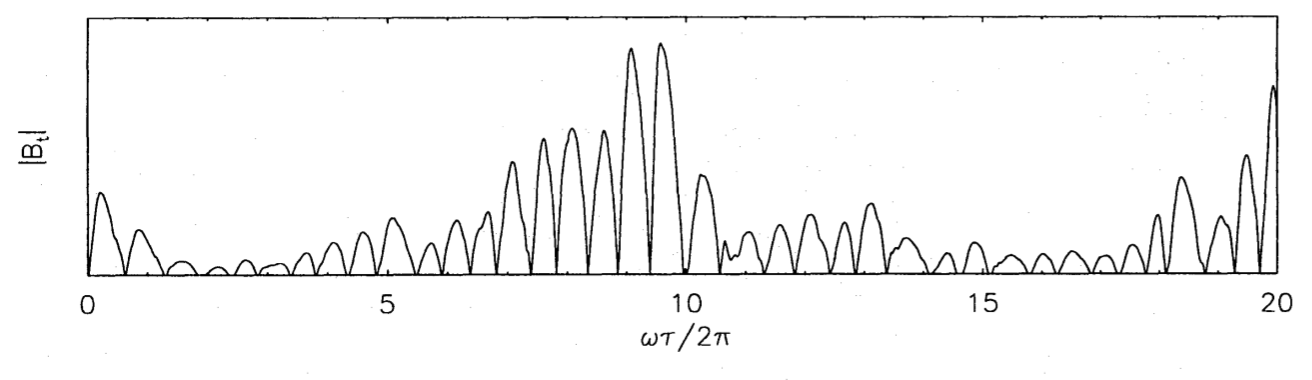}
\caption{A representative case of the solar cycle (as measured by the toroidal field) from a simplified $\alpha$${\rm\Omega}$ dynamo model with stochastic noise in the $\alpha$ effect.  Credit: \citet{OH96}, reproduced with permission $\copyright$ ESO.}
\label{fig:ossen}
\end{figure}

There is a long history studying the modulation of the solar cycle utilizing the stochastic fluctuations in the dynamo model. \citet{C92, H93, OH96, OHS96, GM06, BS08, Mea08} are some examples from a long list of publications in which stochastic fluctuations in the $\alpha$ parameter in their dynamo model were included and found long-term modulations including quiescent period like grand minima in some parameter regimes. 
We would like to 
mention that most of these models also include some nonlinearities, usually the $\alpha$ quenching to stabilize the dynamo. Therefore, it was found that when this $\alpha$ quenching was included, the variability was decreased.  In \Fig{fig:ossen}, 
we present cycles from a simplified mean-field $\alpha{\rm \Omega}$ dynamo model of \citet{OH96} with stochastic fluctuations in the $\alpha$ term. They showed that with a certain amount of fluctuations in $\alpha$, the variability in the modelled cycle closely resembles the variability seen in the observed sunspot data.  
 
\subsubsection{Fluctuations in $\alpha$ effect coupled with dynamic $\alpha$ effect}
\label{sec:mod_coupledalpha}
\begin{figure}
 \centering
\includegraphics[scale=0.40]{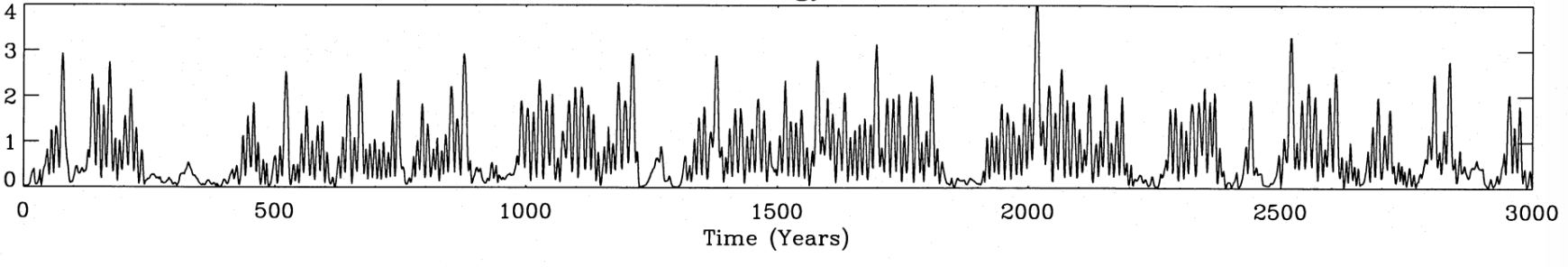}
\caption{Cycle modulations and grand minima (as measured by the magnetic energy) in the dynamo model with stochastic fluctuations in the $\alpha$ effect combined with a (threshold field dependent) dynamic $\alpha$ produced due to the instability in the flux tube at the BCZ.  Reproduced from \citet{Schmitt96} with permission from author.}
\label{fig:alpha_dynamicalpha}
\end{figure}

When the classical $\alpha$ effect is combined with another $\alpha$ effect having a magnetic field-dependent lower threshold, 
a large modulation is expected. The best example for this is the dynamo model coupled with the dynamic $\alpha$-effect which is produced due to the instability in the flux tube at the BCZ \citep{Schmitt85, Chatterjee11}. 
In a mean-field dynamo model, \citet{Schmitt96} included this dynamic $\alpha$ effect in the overshoot layer below the CZ in addition to the classical $\alpha$ effect.  As the dynamic $\alpha$ effect works only when the magnetic field is greater than a threshold field strength, it stops operating when the field falls below this threshold.  \citet{Schmitt96} and \citet{Oss00} showed that when the magnetic field is strong in a normal cycle, both $\alpha$ operate concurrently. However, due to stochastic fluctuations, the magnetic field can occasionally fall below the threshold and the dynamical  $\alpha$ stops operating. This caused the magnetic field to fall drastically---that is the beginning of a grand minimum; see \Fig{fig:alpha_dynamicalpha}. Note that in this case, the magnetic field can suddenly drop to a considerably lower value.   During this quiescent period, the classical $\alpha$ alone slowly grows the field and recovers the model from grand minimum. 

\begin{figure}
\centering
\includegraphics[scale=0.3]{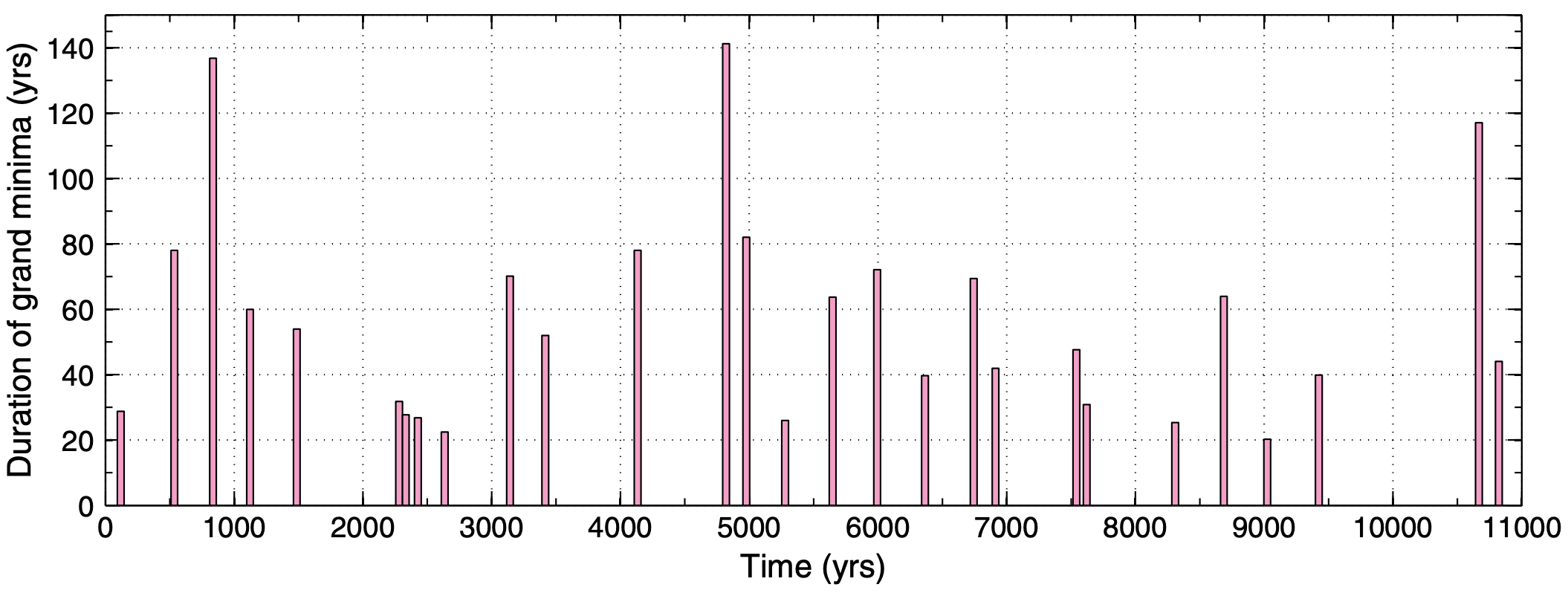}
\caption{The durations vs the times of their occurrence of the grand minima in the 2D \bl\ dynamo model of \citet{CK12}.}
\label{fig:CK_gm}
\end{figure}

\subsubsection{Fluctuations in \bl\ process}

\begin{figure}
\centering
\includegraphics[scale=0.22]{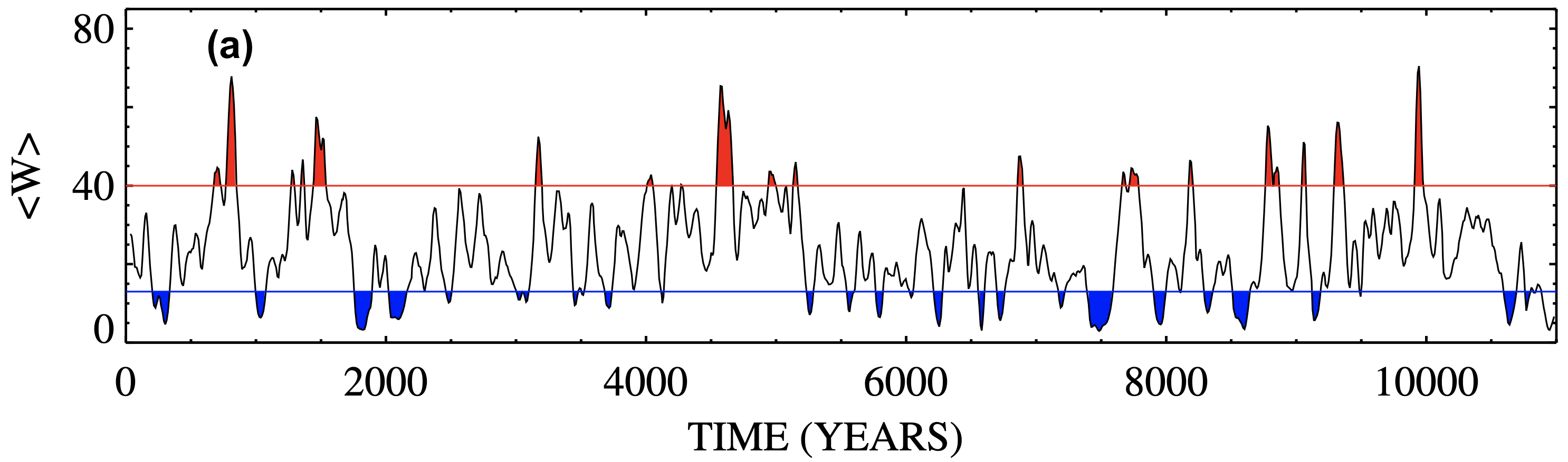}
\includegraphics[scale=0.15]{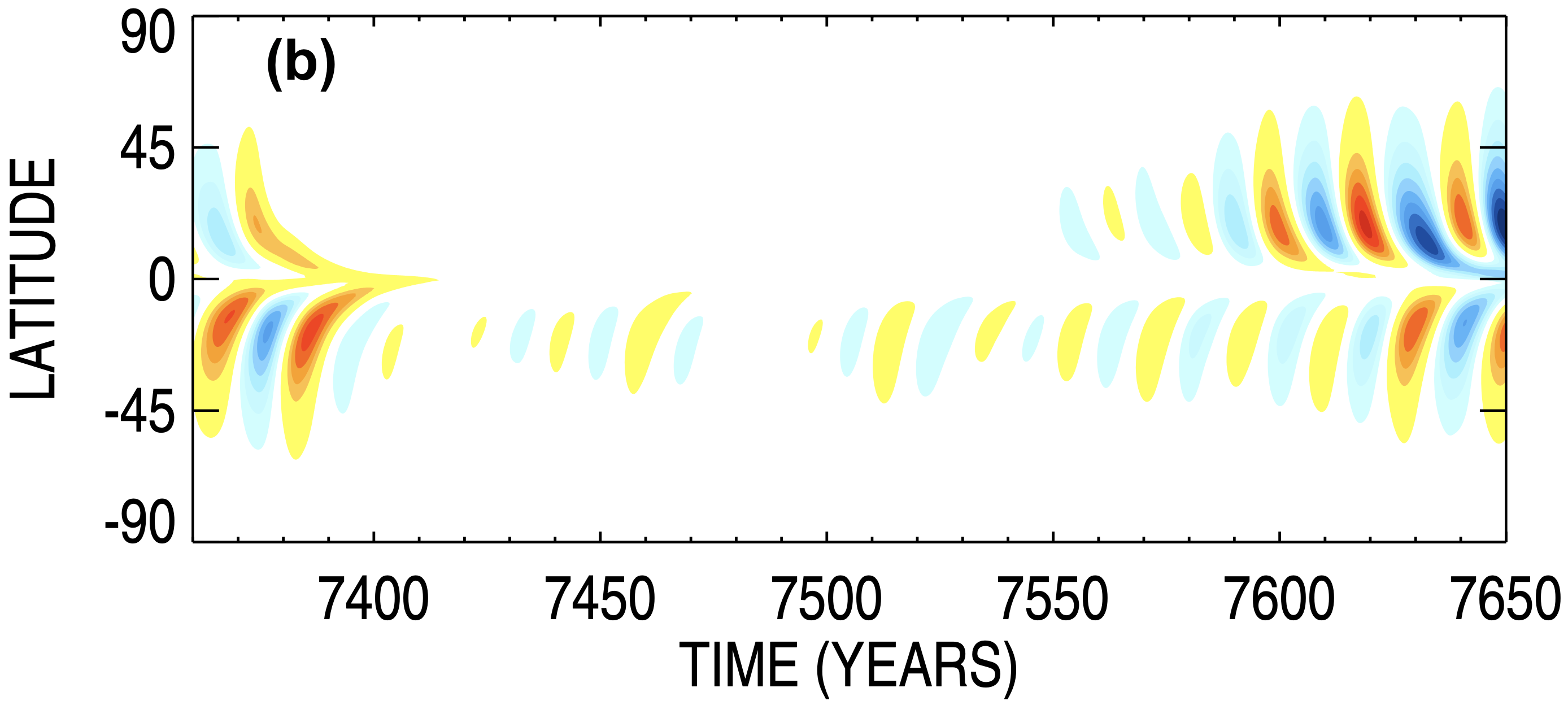}
\caption{(a) A proxy of the smoothed sunspot number from the dynamo model of \citet{OK13}. Blue and red shaded regions correspond to the grand minima and maxima \citep[using the same definition as used in][]{USK07}. (b) Toroidal field as a function of latitude and time, highlighting a grand minimum. Reproduced from \citet{OK13} with permission.}
\label{fig:OKgm}
\end{figure}

 For about the last two decades, \bl\ type flux transport dynamo models have been extensively used to explain the variabilities in the solar cycle. The first landmark paper in this series came from \citet{CD00} who included stochastic fluctuations in their 2D (axisymmetric) flux transport dynamo model. For this, they added a stochastic term with a coherence time of a month in the $\alpha$ parameter, the \bl\ source term in their axisymmetric model. They essentially replaced $\alpha$ by $\left[1+ s~ \sigma(t)\right] \alpha$ in \Eq{eq:pol}, where $\sigma$ is random deviate within $[-1,1]$ whose value is updated every month and $s$ determines the level of fluctuations.
 They found some modulation in the solar cycle, including the observed weak anti-correlation between the cycle duration and the amplitude with $200\%$ fluctuations ($s=1$) in $\alpha$. Later \citet{Cha07} showed that the long time delay inbuilt in this type of \bl\ dynamo model naturally reproduces a \go\ like pattern in
 the modelled solar cycle (more in \Secs{sec:timedelaymod}{sec:origin_OErule}). 
 Fluctuations in the \bl\ process can also lead to a large variation in the poloidal field with occasional dips in the polar field (as observed in the solar magnetic field), which can lead to double peaks (Gnevyshev gaps) and spikes in the following cycle \citep{KMB18}. Large fluctuations in the \bl\ process also lead to Maunder-like grand minimum as shown initially by \citet{Cha04} and later by many other authors \citep{CK09, Passos12, Ha14, Pas14} in different \bl\ type dynamo models. In particular, \citet{CK12} estimated the amount of variation in the polar field at the end of a cycle (a cumulative effect of the fluctuations in the \bl\ process) and the variation in the \mc\ based on indirect observations and included those into their  
 high diffusivity dynamo model. They found the correct frequency of the grand minima in the last 11,000 years (\Fig{fig:CK_gm}). Another work was by \citet{OK13} who also made an estimate of the level of fluctuations 
 in the \bl\ process by computing the contribution 
to the 
polar field from the sunspot group data of Royal Greenwich, Kodaikanal and Mount Wilson Observatories.
 They found that the statistic of grand minima are consistent with the Poisson random process, which indicates that the initiation of grand minima is
 independent of the history of the past minima \citep[also see][]{KC13}. They also showed that there is a correlation between the occurrence of grand minima and the deviation from the dipolar parity and thus the hemispheric asymmetry; also see \citet{Nagy17, HN19} for the same conclusion from  different \bl\ models. \Fig{fig:OKgm}(a) shows the smoothed \citep[in the same way as done in][]{USK07} sunspot number from an 11000-year long simulation done by \citet{OK13} in which the red and blue shaded areas represent the grand maxima and minima, respectively.
 
  The hemispheric asymmetry which is a robust feature during grand minima is also reflected in a typical grand minimum as presented in \Fig{fig:OKgm}(b).   
 The fluctuations in the \bl\ process 
of north and south hemispheres are uncorrelated 
and thus hemispheric asymmetry is unavoidable during grand minima in this model and 
 also other dynamo models with 
 fluctuations in the \bl\ process \citep{Pas14, KM18}. The hemispheric asymmetry, however, cannot remain for multiple cycles as the diffusive coupling at the equator 
tends to 
smooth out the asymmetry acquired due to fluctuations \citep{CC06, KM17}. Another way that north-south asymmetry 
 in the magnetic field can come about in these models is due to the random excitation of the quadrupolar mode by the stochastic fluctuations in the \bl\ process \citep{SC18}. 
 
Further support for the stochastic origin of the long-term modulation of the solar cycle came from \citet{CS17} who studied the following `stochastic' normal form model.
 \begin{equation}
 {d X } =   \left(\beta + i \omega_0 -   (\gamma_r + i \gamma_i) |X|^2\right) X {d t} + \sigma X dW_c = 0,
 \end{equation}
 where $X$ is a complex quantity whose real and imaginary components give the toroidal and poloidal fields, $\beta$ determines the growth rate of the dynamo and thus the supercriticality,  $\omega_0$ sets the magnetic cycle frequency, $\gamma_r$ and $\gamma_i$ regulate the nonlinearity of the model and determine the cycle amplitude, $W_c$ represents a complex Wiener process and $\sigma$ is a measure of added noise. The values of all these parameters are fixed by the observations. \citet{CS17} showed that in the weakly nonlinear regime, the variability of the solar cycle as seen in the reconstructed data over the past 9000 years can be modelled using this stochastic normal-form model \citep{CS19}. They have also tested this idea using a 1D \bl\ type dynamo model constrained by observations.

While most of the \bl\ models are kinematic \citep[see][for an exception to this]{BC22}, recently \citet{Fadil17} utilized the 2D nonkinematic 
dynamo model of \citet{Remp06} to study the nature of the grand minima and maxima. 
In this work, they considered random fluctuations in the angular momentum transport process in addition to the \bl\ term. This caused some nonlinear interaction between the flow and the fields. 
Even in this model, they found that the occurrences of grand minima and maxima are 
largely described by memoryless processes.
In this model, it is also expected to observe modulation in the flow. They found that 
the radial differential rotation tends to be larger during grand maxima, while it is smaller during grand minima. 
The latitudinal differential rotation, on the other hand, is found to be larger during grand minima 
in agreement with the data by \citet{RN93}.
The meridional circulation speed tends to be faster during grand minima.

\begin{figure}
\centering
\includegraphics[width=1.0\columnwidth]{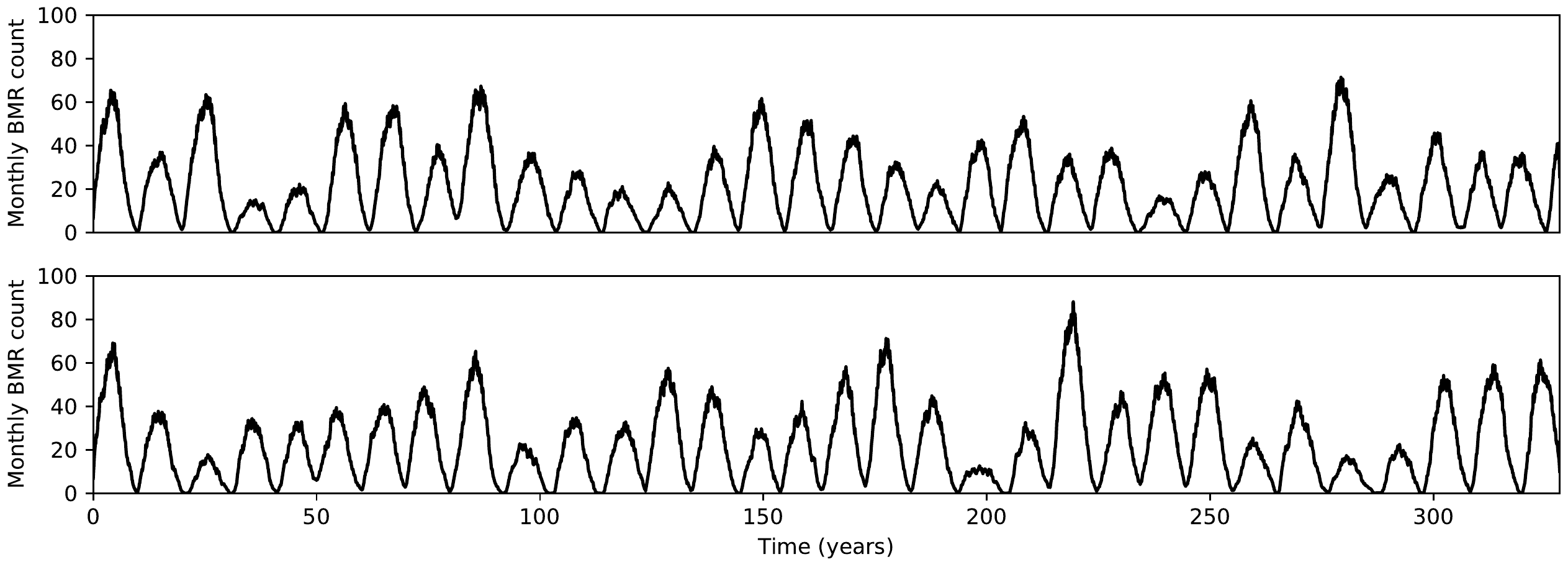}
\caption{Time series of the monthly BMR number (pseudo-SSN)
from the simulation with observed tilt scatter of the $ 2\times2$D kinematic \bl\ model \citep{LC17}. 
Panels are obtained from two different realizations of dynamo simulations at same parameters. 
}
\label{fig:LC17}
\end{figure}

\begin{figure}
\centering
\includegraphics[width=1.0\columnwidth]{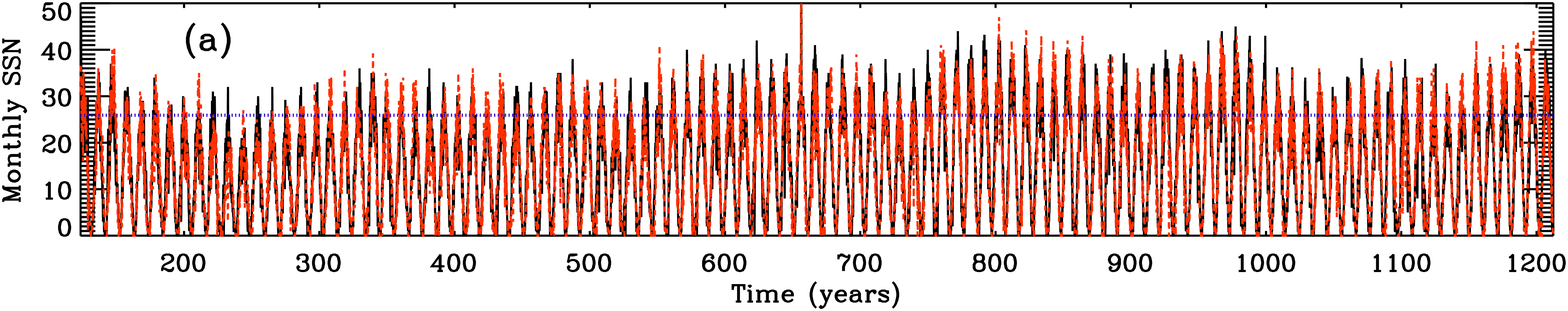}
\includegraphics[width=1.0\columnwidth]{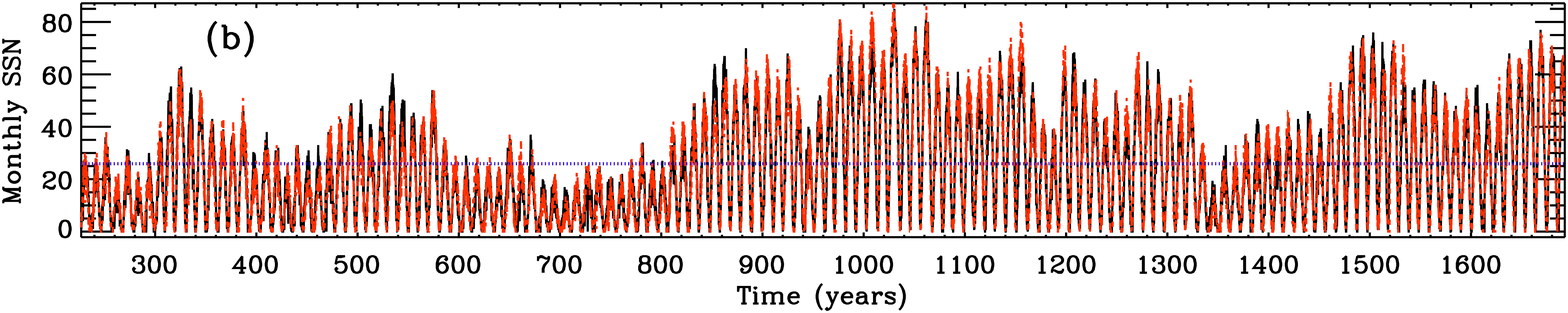}
\includegraphics[width=1.0\columnwidth]{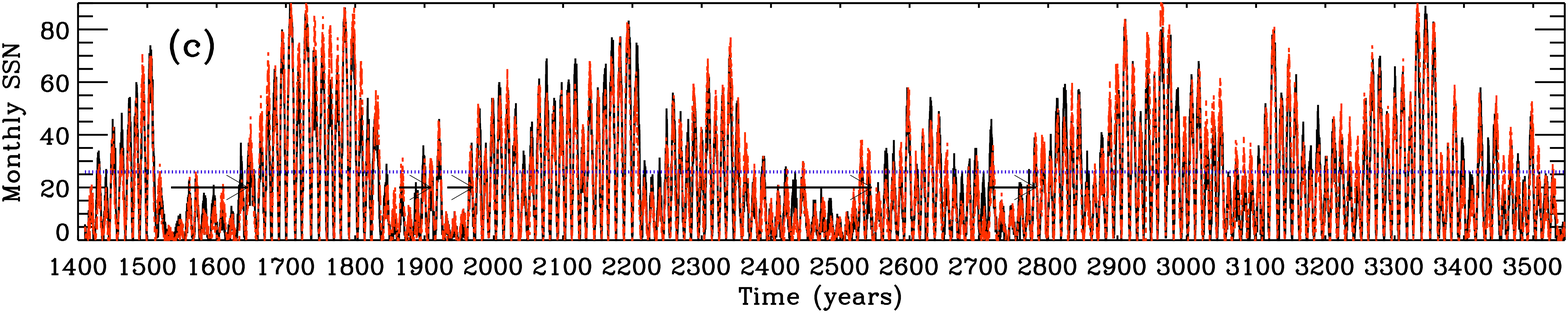}
\caption{Time series of the monthly BMR number
from a 3D kinematic \bl\ dynamo model (a) without tilt scatter, (b) with a Gaussian scatter of $\sigma_\delta=15^\circ$ (close to the observed value), and (c) $\sigma_\delta=30^\circ$, respectively taken from Runs~B9, B10, and B11 of \citet{KM17}.
The horizontal line shows the mean of peaks of the monthly group numbers obtained for last $13$
observed solar cycles.
Arrows in (c) represent the locations of grand minima.}
\label{fig:KM17}
\end{figure}

 Recently, two comprehensive kinematic \bl\ dynamo models, namely, $ 2\times2$D model \citep{LC17, Nagy17} 
 and 3D model \citep{MT16,KM17} were used to model the cycle modulations. 
 The good thing about these models is that the BMRs are explicitly deposited in these models and thus all the observed (statistical) properties of the BMRs are captured by and large.  \citet{LC17, KM17} showed that the 
 randomness associated with the BMR production are the major causes for the long-term modulation in the solar cycle.
 Both these groups included the observed scatter around Joy's law tilt (using a Gaussian distribution) and found considerable amount of variation, including north-south asymmetry and grand minima in the solar cycle (\Figs{fig:LC17}{fig:KM17}). 
 As the level of fluctuations is increased, the variability and the number of grand minima is increased (\Fig{fig:KM17}). 
 We note that the observed distribution in the delay time of the BMR eruption and the BMR flux also give some modulation 
 in the solar cycle as seen in \Fig{fig:KM17}(a) in which there is no scatter imposed around Joy's law.
 
\subsubsection{Does the \bl\ process operate during grand minima?}
\label{doesBLoperate}
Dynamo models including only the \bl\ process for the poloidal source is so successful in reproducing the observed 
features of the solar cycle, it is natural to ask the question whether the \bl\ process operates during \gm.

\begin{figure}
\centering
\includegraphics[width=1.0\columnwidth]{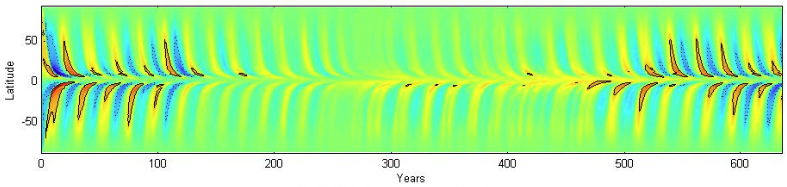}
\caption{Grand minima reproduced from 2D flux transport dynamo model with additional 
$\alpha$ effect operating in the bulk of the CZ \citep{Pas14}. Colour shows the toroidal field at the BCZ and the contours show the areas where the field exceeds the threshold for the model-spot eruptions. Note that in this model, the recovery from the grand minimum is due to the 
$\alpha$ effect; see also \cite{olc19}. 
Image reproduced with permission from authors.
}
\label{fig:passos}
\end{figure}

To operate the \bl\ process, we need ``tilted'' BMR. However, observations, particularly the early ones, found only a few spots during the \mm. Thus, one would expect that the \bl\ process will not operate, or be inefficient during the \mm\ and possibly during other grand minima. The classical (helical) $\alpha$ \citep{Pa55} which efficiently operates in the sub-equipartition field is a obvious candidate for the generation of the poloidal field during these episodes \citep{KC13, Ha14}. 
\citet{Pas14} performed simulations of 2D \bl\ dynamo model with a weak classical $\alpha$ operating in the whole CZ. They showed that this additional $\alpha$ produces poloidal field and recovers the model from \gm\
when the \bl\ process stops operating (\Fig{fig:passos}); also see \citet{olc19} for another beautiful demonstration of this idea in 2D$\times$2D model with explicit BMR deposition.  
Unlike most of the \bl\ dynamo models\footnote{Even the models of grand minima by \citet{Kar10, CK12, KC13} using the Surya code \citep{CNC04} although includes a lower threshold for the spot eruption, the \bl\ process still operates in these models because some toroidal field rises to the upper layer due to upward meridional flow and diffusion.}, in these models, the \bl\ process is stopped operating at low field regime by introducing a lower threshold
and thus only helical $\alpha$ operates during \gm.  
The mechanism of this type of dual dynamo model is similar to the model of \citet{OH96} in which a classical $\alpha$ and the dynamical $\alpha$ driven by the magnetic buoyancy were incorporated (\Sec{sec:mod_coupledalpha}).

In any rotating convective layer (like the solar CZ), generation of $\alpha$ effect is natural. However, its nature and how strong its value in the CZ is still uncertain. Thus introducing this $\alpha$ in the model, brings several unknown parameters. On the other hand, the following facts support the operation of \bl\ process during \mm\ and other \gm.
(i)  \mm\ was not completely devoid of sunspots. 
 Recent analyses clearly show that some spots were observed during the \mm\ \citep{Zolotova_2015, Uso15, Vaq15, ZP16}.
 (ii) Even a few BMRs can produce an appreciable amount of poloidal field which, if not decayed considerably, can 
 slowly produce enough toroidal field and recover the Sun from \gm\ phase. In fact, \citet{Ca12} showed that  in the Sun 
 the diffusion of the poloidal field through the surface is negligible---this in the flux transport dynamo models
 can be achieved by including a downward magnetic pumping \citep{KC16}. 
 (iii) Smaller BMRs (including ephemeral regions) produce little contrast in the white light and were not detectable as spots with the telescopes of the Maunder minimum epoch
 \citep{Jha20} but their eruption rate is large \citep[smaller the BMR, larger is the emergence rate;][]{HST03}.
 These small BMRs have some non-zero tilt and they can produce some polar field during \gm\ \citep{SK12, Tla13, Jha20}.

\begin{figure}
\centering
\includegraphics[width=1.0\columnwidth]{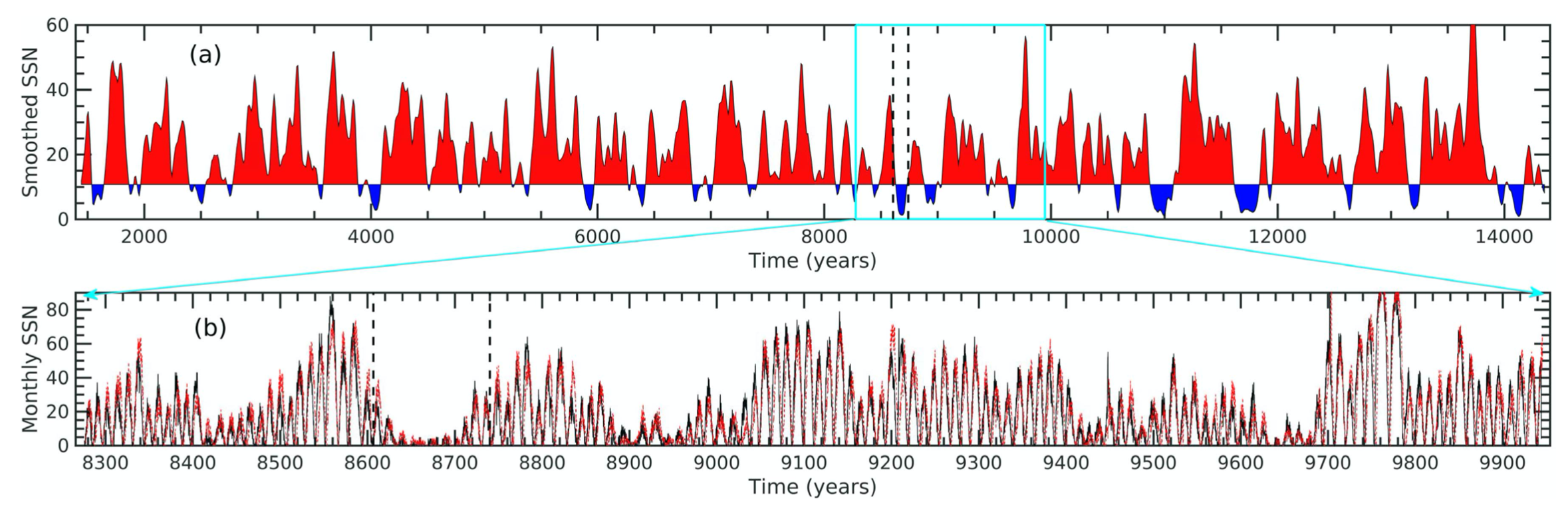}
\caption{(a) Decadal-binned and smoothed BMR number from the model of \citet{KM18} 
with randomness in the BMR properties (mainly scatter around Joy's law tilt).
(b) Shows the zoomed-in portion of the monthly smoothed BMR number. 
}
\label{fig:KM18}
\end{figure}

\begin{figure}
\centering
\includegraphics[scale=0.25]{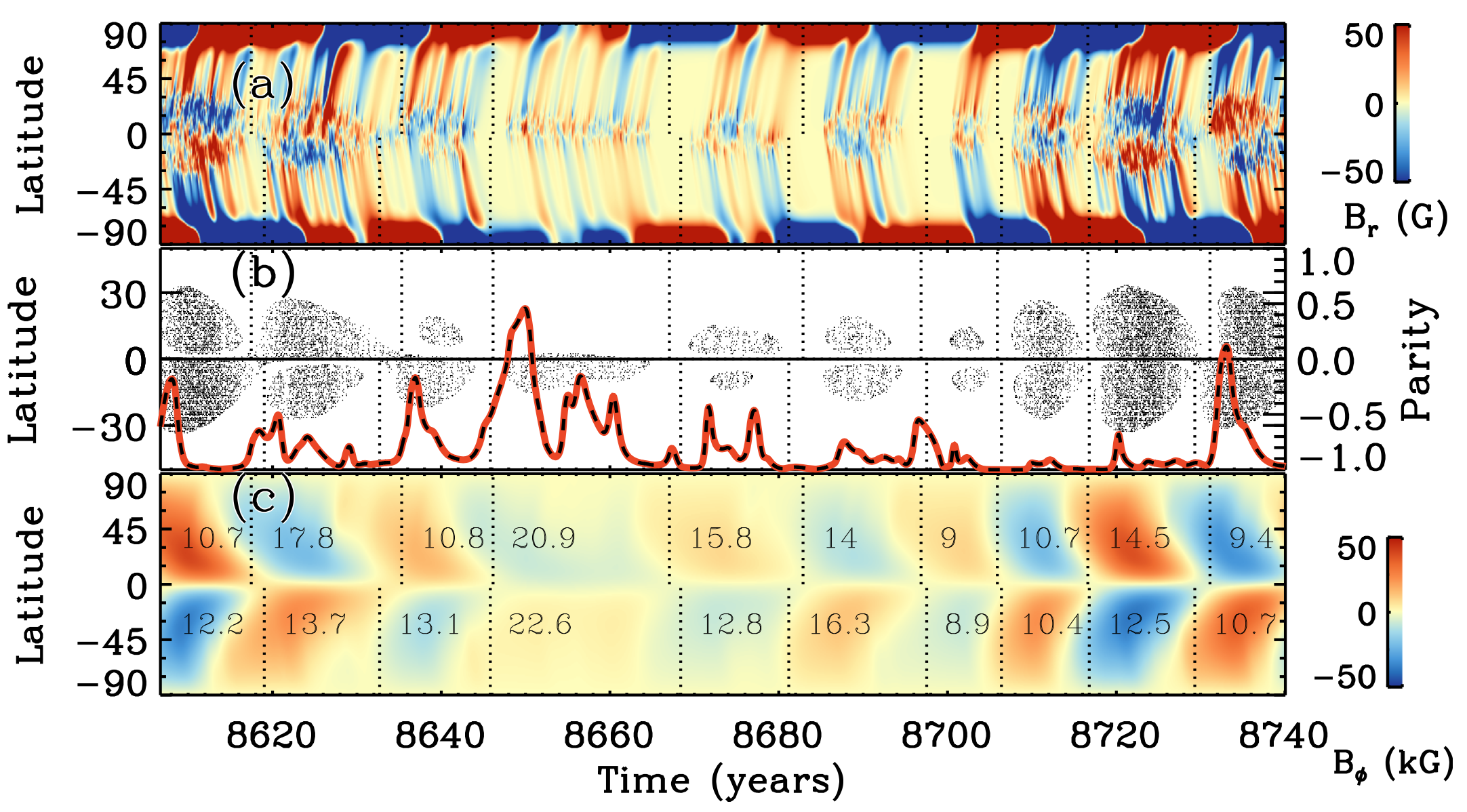}
\caption{Evolution of the (a) surface radial field (b) BMR eruptions and the hemispheric parity of the toroidal field (dashed line) and (c) the toroidal field at BCZ from a grand minimum presented in \Fig{fig:KM18}(b) (marked by dashed lines); also see \citet{KM18}. 
}
\label{fig:KM18b}
\end{figure}

By including a downward magnetic pumping, \citet{KM18} showed that the model with stochastic properties (tilt scatter) in BMRs can recover 
from grand minima without any additional source for the generation of the poloidal field; see \Fig{fig:KM18}. They found that during grand minima as the poloidal magnetic field does not decay (due to pumping), it keeps on supplying the toroidal field and thus the model continues to produce BMR at a low rate even during grand minima (\Fig{fig:KM18b}). The poloidal field generated from these few BMRs is alone sufficient to recover the model to the normal phase. 
Their model reproduces most of the features of the grand minima (including frequency of grand minima, longer cycles and strong hemispheric asymmetry during grand minima).
  

\subsubsection{Variability vs dynamo supercriticality}
\label{sec:supercriticality}
\begin{figure}
\centering
\includegraphics[scale=0.40]{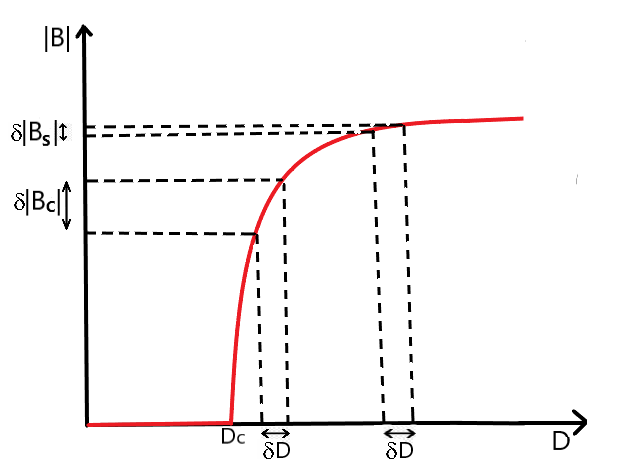}
\caption{A Hopf bifurcation diagram, showing the transition from a fixed point to dynamo instability. 
This is a typical variation of the magnetic field strength ($|B|$) vs dynamo number ($D$) in dynamo model with any nonlinear quenching mechanism as long as $D$ is not much larger than $D_c$.  
Here, $D_{\rm c}$ is the critical $D$. $\delta |B_c|$  and 
$\delta |B_s|$ are the amplitude variations of the magnetic field for a given change in $D$ in two different regimes of the dynamo.
}
\label{fig:BvsD}
\end{figure}

Whenever there is any change in the dynamo number 
$D$ ($= \alpha_0 {\rm \Delta {\rm \Omega}} R_\odot^3 /\eta_0^2$, where $ \alpha_0$ is the strength of $\alpha$ effect, ${\rm\Delta} {\rm \Omega}$ is the amount of shear in the CZ, and $\eta_0$ is the diffusivity), there will be a change in the amplitude of the magnetic field. Thus, the cycle modulation due to fluctuations in the dynamo parameter is obvious.  
However, for a given level of fluctuations and the form of nonlinearity, the amount of variability 
depends on the value of $D$ or the regime operation of the dynamo. This is apparently seen in \Fig{fig:BvsD} that the same amount of variation in $D$ causes a large variation in magnetic field 
when the dynamo operates near the critical transition ($\delta |B_c|$) and a small variation when the dynamo operates in 
supercritical regime ($\delta |B_s|$)
\footnote{Another way to understand this result is from the dynamo instability. In a slightly supercritical regime, the amplitude B of magnetic cycles follows a general rule $B \propto (D - D_{c})^{1/2}$ for all instabilities \citep{LL87}. Thus, $\delta B / \delta D$ decreases with the supercriticality.}. 
The reason for this is not difficult to understand. 
When the dynamo operates near the critical transition, a small $D$ makes the growth rate of the magnetic field small and the dynamo weakly nonlinear. Now consider a scenario when the magnetic field 
has become weak due to a reduction of $D$ (or $\alpha_0$) and after some time
due to fluctuations, $D$ has increased. Then the dynamo will (almost linearly) amplify the field for a long time before the nonlinearity becomes important and thus the net growth of the field will be large. On the other hand, if the dynamo operates in a highly supercritical regime, then the nonlinearity will quickly suppress the dynamo growth and the net amplification of the field will be small.

\begin{figure}
\centering
\includegraphics[scale=0.25]{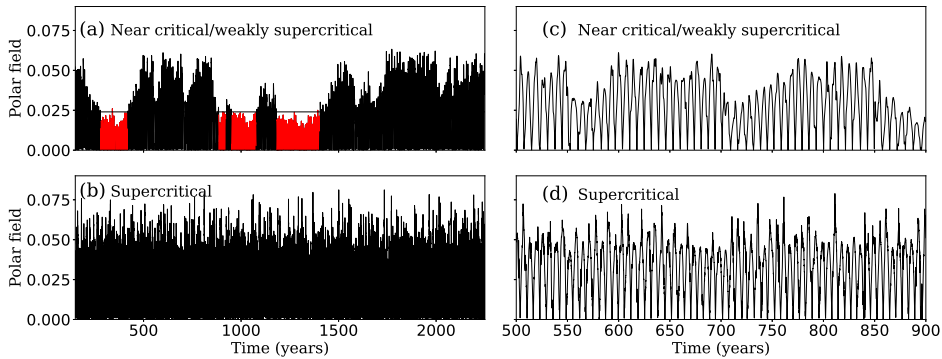}
\caption{Polar field (averaged over $55^\circ$ latitude to north pole) from a dynamo simulation in which the dynamo operates (a) near-critical or weakly-supercritical regime ($D/D_c = 2$) and (b) supercritical regime ($D/D_c = 6$). (c) and (d) are showing the cycles for 400~years from the long data shown in (a) and (b), respectively. 
The red portions represent extended weaker activity (the grand minima). The figures 
are produced from the model presented in \citet{kumar21b}.
}
\label{fig:crit_supercrit}
\end{figure}

The above discussion also means that in the near-critical (or weakly supercritical) regime, we expect long-term modulation in the cycle and extended grand minima. In this regime when the field becomes weak due to fluctuations, the dynamo will take a long time (several cycles) to grow the field and this will tend to produce a smooth long-term variation. In contrast, in the super-critical regime, we do not expect much long-term modulation in the cycle amplitude and no extended grand minima because when the magnetic field falls to a low value, the dynamo will quickly increase the field in a cycle. This is clearly seen in \Fig{fig:crit_supercrit}; also see \citet{Vindya21,tripathi21,albert21}. Furthermore, for the given diffusive and advective transports, 
the long-term memory of the field should depend on the supercriticality of the dynamo. 
\citet{kumar21b} showed that in the weakly supercritical dynamo, the long-term memory 
of the polar field persists for multiple cycles. However, when the supercriticality is increased, 
the multi-cycle memory is reduced to only one cycle. 
In fact, for rapidly rotating young stars, we expect the dynamo to be strong (convective motion is more helical) and thus we do not expect extended grand minima there \citep{V22, V23}. This is congruous with the stellar observations that the grand minima are detected only in the slowly rotating stars \citep{BoroSaikia18, Olah16, Shah18, garg19, Baum22}. 

The above discussion, although seems to be promising, has a subtlety. 
In the highly supercritical regime, when the dynamo number is increased to a very high value, 
the variability of the magnetic field may not remain small because the dynamo may enter into a more complex regime. In this regime, dynamo modes with different base periods can emerge and in turn the 
linear superposition and nonlinear coupling of different modes is expected to
introduce an increasingly strong and complex modulation of the periodic behaviour \citep{Cha07, SC18,albert21}.  
In addition, even without the presence of other modes and other base periodicities, the
iterative map of \citet{Cha01} and the flux transport dynamo model inbuilt with a time delay and nonlinearity \citep{CSZ05}
enter into chaotic solution through sequence of period-doubling bifurcations. 
However, we must remember that to enter the dynamo into this region, the $D$ 
must be much larger than $D_c$; see e.g., Fig. 3 of \citet{CSZ05}; also see the discussion given at the end of \Sec{sec:causes_nonlinear}. Furthermore, we find that \bl\ dynamos with the popular 
$\alpha$-quenching of the form: $1/[ 1 + (B / B_0)^2] $ does not lead to chaotic solution;
the solution remains stable all the way to a very large value of $D$. 


It remains a big question that what is the supercriticality of the Sun. Stellar observations indicate that probably our Sun is not too supercritical \citep{Met16, KN17}. 
Furthermore, as only the weakly supercritical dynamo produces grand minima \citep{Vindya21, kumar21b, CS17} and somewhat smooth cycle variation and Sun does produce grand minima, we expect that the solar dynamo is probably not operating in highly supercritical regime.


 \subsection{Specific nonlinearities in the \bl\ process}
 \label{sec:BLnonlinear}
 In the \bl\ solar dynamo also, the magnetic field acts on the flow and gives a nonlinearity in the model. However, given the fact that the observed differential rotation has only a little variation over the solar cycle, the poloidal to toroidal 
field conversion,
i.e., the ${\rm\Omega}$ effect, is largely linear. On the other hand, the toroidal to poloidal part of the \bl\ models is not due to the classical $\alpha$ effect which experiences a catastrophic quenching due to magnetic helicity conservation \citep[Sec. 8.7 of][]{BS05}, rather it is due to the \bl\ process. The latter is a nonlocal process and does not experience catastrophic quenching \citep{KO11c}.  However, due to a lack of understanding in the past, most of the \bl\ dynamo models included a simple quenching of the form
 $1 / \left(1 + (B / B_0)^2 \right)$ in the poloidal field term to limit the growth of magnetic field \citep{Cha10}.
 Fortunately, in recent years some potential candidates for nonlinearity have been identified which we discuss below. 
 
\begin{itemize}
\item{\bf{Tilt quenching}}
\end{itemize}

The tilt angle of BMR plays a crucial role in generating the poloidal field in the Sun. Theory based on the thin flux tube approximation suggests that the tilt is produced due to the torque induced by the Coriolis force acting on the east-west flow emerging from the apex of the rising flux tube of toroidal 
field \citep[][also see Sec.\ 5 of \citet{Fa21}]{DC93, FFM94}. 
If a flux tube has a strong magnetic field, then there will be two consequences. One is that the flux tube will rise quickly due to strong magnetic buoyancy and thus the Coriolis force will not get much time to induce 
the tilt. The other consequence is that the flux tube will have strong magnetic tension which will oppose the torque. Both of these effects will reduce the tilt. Based on this theoretical concept, we expect the tilt to decrease with the magnetic field in the BMR forming flux tube. This, the so-called tilt quenching may be a potential source for the nonlinearity in the \bl\ type dynamo models.

 \begin{figure}
\centering
\includegraphics[width=0.75\columnwidth]{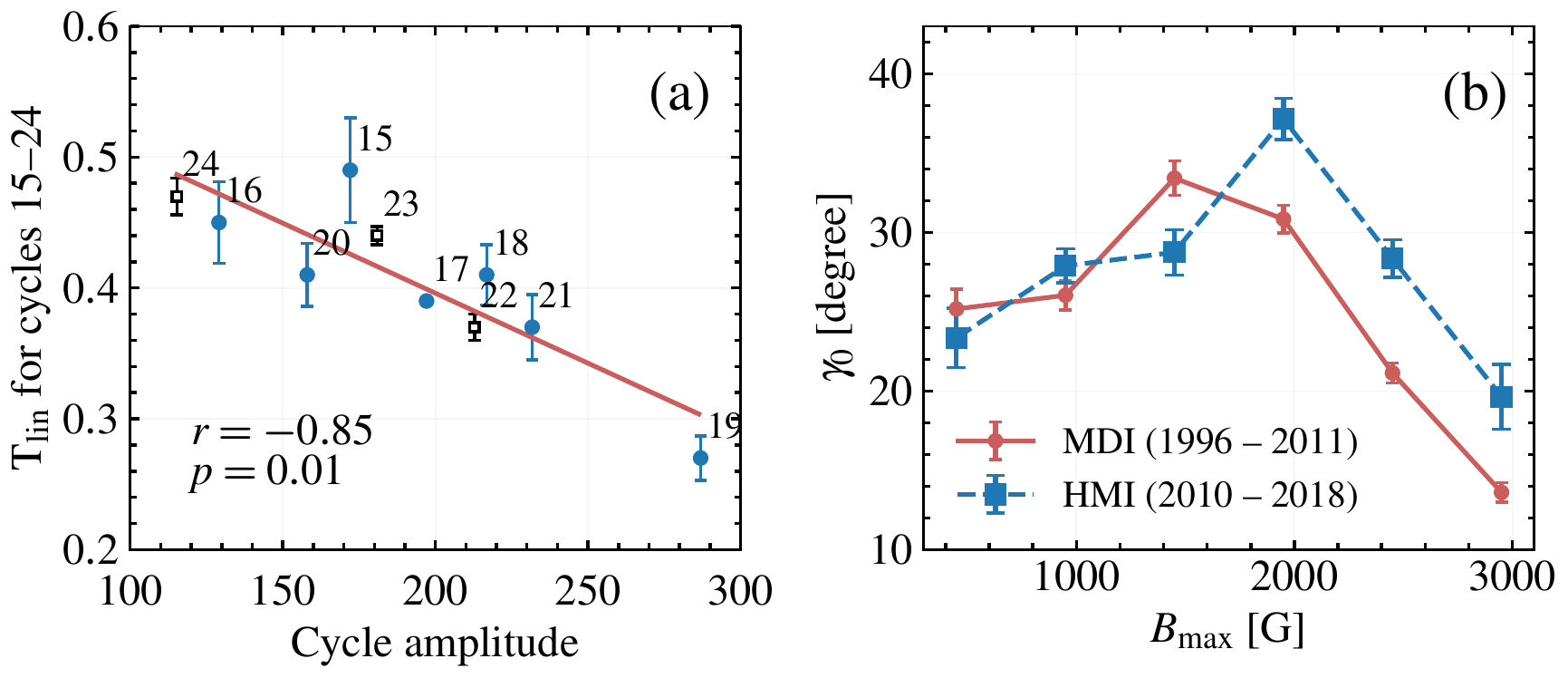}
\caption{(a) Tilt coefficient as computed from the mean tilt normalized by the mean latitude vs the sunspot cycle amplitude \citep{Jiao21}; also see \citet{Das10}. (b) Slope of Joy's law as a function of the maximum field strength of the BMR \citep{Jha20}. }
\label{fig:tilt_quench}
\end{figure}

The observational support for the tilt quenching however is limited. \citet{Das10} have found an anti-correlation between the cycle amplitude and the cycle-average tilt angle of the sunspot group normalized by the mean latitude from the white-light data of Mount Wilson and Kodaikanal 
Solar Observatories \citep[also see][for the corrected plot for Mount Wilson data]{Das13}. However, other studies do not find a statistically significant relationship for this \citep[e.g.,][]{Wang15}. Also, studies have shown that the tilt angle measured from the white light data can be significantly different than that obtained from the magnetic field data \citep{Poisson20}. \citet{Jiao21} carefully examined
the previous methods of estimating tilt angles from Kodaikanal and Mount Wilson, supplemented by 
tilt angles from Debrecen Photoheliographic data and show that the tilt is statistically 
anti-correlated with the cycle strength as shown in \Fig{fig:tilt_quench}(a).

While the above studies explored the variation of the cycle-average tilt with the cycle strength, \citet{Jha20} 
examined the tilt of BMR within the cycle from the line-of-sight magnetograms of Michelson Doppler Imager (MDI onboard Heliospheric Observatory (SOHO); 2010--2018) and Helioseismic and Magnetic Imager (HMI onboard Solar Dynamic Observatory (SDO); during 1996--2011). They showed that the BMR tilt has non-monotonous dependence on the BMR's field strength; in the small field regime, the tilt increases and in the large field regime it decreases. This is shown in \Fig{fig:tilt_quench}(b). 
We note that in this study, the data were used from Cycles 23 and 24 which are 
weak cycles and for these very little reduction of tilt was seen in the study of \citet{Jiao21}. 
Therefore, the limited data used in \citet{Jha20} could not predict the exact magnetic field-dependent 
form of the BMR tilt.

 \begin{itemize}
\item{\bf{Flux loss due to magnetic buoyancy}}
\end{itemize}

Magnetic buoyancy is the key for the formation of sunspots or more generally BMRs \citep{Pa55b}. Models under thin flux
tube approximation showed that when a portion of the toroidal flux tube with sufficient field strength at the BCZ becomes magnetically buoyant, it rises to surface to give rise to a BMR. In this process, the portion of the flux tube from where the flux is depleted becomes inefficient for flux eruption for some time. 
As the flux emergence 
happens only when the field strength exceeds a certain value and after each flux emergence the flux is reduced locally, this introduces a nonlinearity in the dynamo. This nonlinear loss of toroidal flux 
plays an important role in limiting the growth of the magnetic field in the Sun \citep{SS89, NC00, CNC04}. 
Studies have shown that dynamo models including magnetic buoyancy reproduces  observations better
\citep{SS89, Haz15}.
  
\begin{figure}
\centering
\includegraphics[scale=0.32]{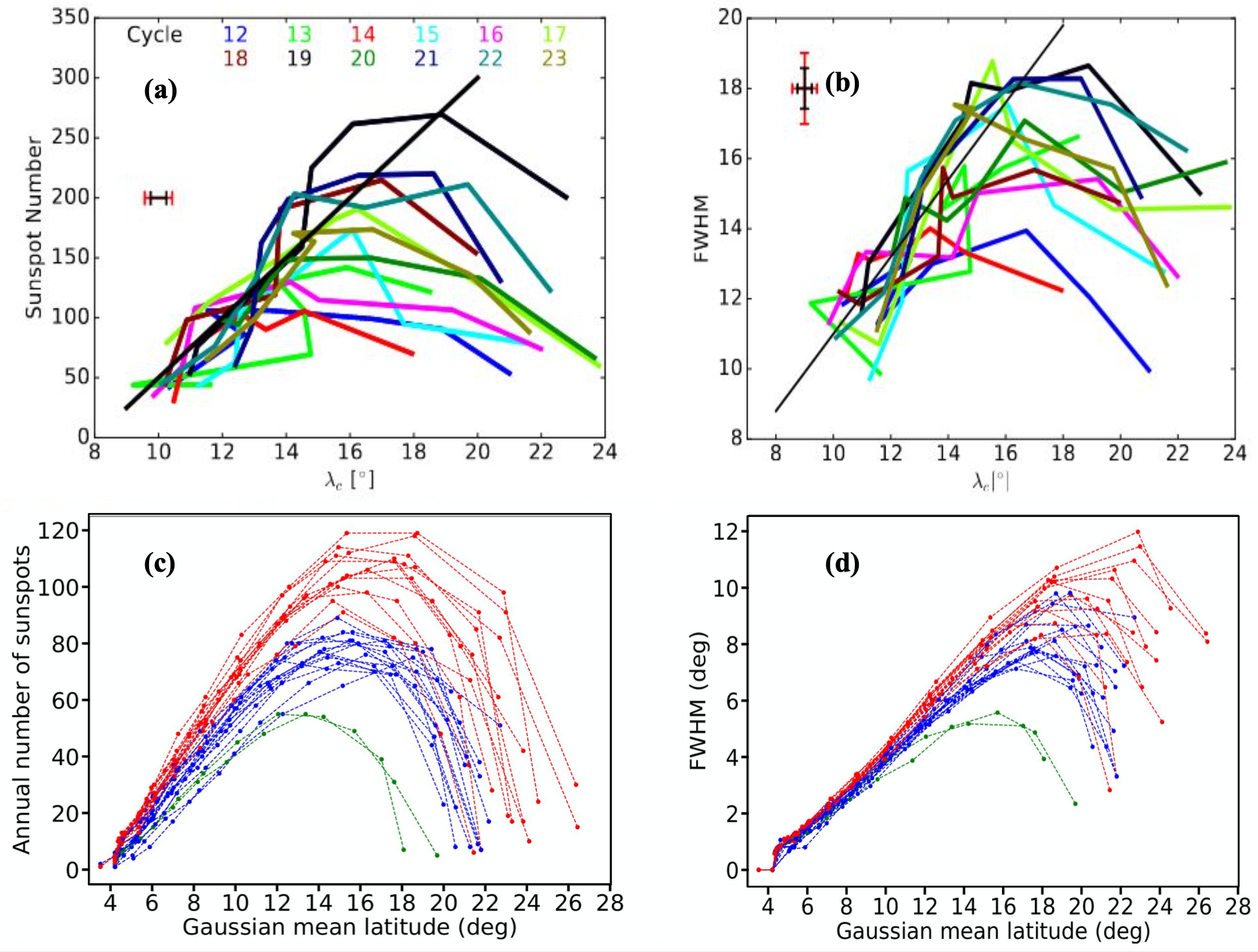}
\caption{The trajectory of the cycles when plotted in terms of annual sunspot number (a,c) and full width at half maximum (FWHM) of the latitudinal distribution (b,d) against the central latitude of the annual distribution. (a,b) From observations \citep{CS16} (with permission from authors). (c,d) From dynamo simulation with buoyancy-induced toroidal flux loss \citep{BKC22}. 
The plots show that the beginning phases (right part of the curves) of the cycles are widely different depending on their strength yet the decline phase are quite similar irrespective of their strength. Also see \citet{Talafha22} for modeling these features using a different model.
}
\label{fig:CS16}
\end{figure}

By introducing the magnetic buoyancy in a simple way,  \citet{BKC22} explained the observed latitudinal variation of the solar activity over the cycle as shown in \citet{W55, CS16} also see \Fig{fig:CS16} (a-b). 
Let us explain how the flux loss helped to explain the observed feature. In the \bl\ dynamo model, the poloidal magnetic field largely produces a toroidal field in the higher latitudes of the deeper CZ and then this toroidal field is advected towards the lower latitudes. 
During its journey, when the toroidal field exceeds a certain value $B_c$ (the threshold for BMR eruption), 
it starts producing BMRs at higher latitudes. As the ${\rm\Omega}$ effect keeps on producing the toroidal field, 
the activity grows while the toroidal flux is advected towards the equator. 
Now consider a cycle becomes strong. The strong cycle  
 starts producing BMRs at high latitudes and the activity (number of BMR) rises rapidly (strong cycles rise rapidly).
 Rapid growth means the toroidal flux loss occurs at a faster rate. Quickly the magnetic field at BCZ 
 becomes comparable to $B_c$ and the activity does not grow further.  
Any further generation of the magnetic field will then be compensated by the flux loss due to BMR emergence and the cycle decline at the same rate for all cycles; \Fig{fig:CS16} (c-d).  That is why in strong cycles the activity begins to decline when the activity belt is already at higher latitudes.

We would also like to mention that without invoking the flux loss due to magnetic buoyancy, only the cross-equatorial diffusion was used to explain 
this universal decline of the solar activity in different studies \citep{CS16, Talafha22}.

\begin{itemize}
\item{\bf{Latitudinal quenching}}
\end{itemize}

The observational facts presented in \Fig{fig:CS16}(a-b) also suggest that strong cycles on average
produce BMRs at high latitudes and vice versa \citep[also see][]{W55, SWS08, MKB17}. On the other hand,
we know that when BMRs appear at higher latitudes, they are less efficient in generating a poloidal field due to poor cross-equatorial cancellation \citep{JCS14, KM18, Pet20}. In contrast, when BMRs appear near the equator, it becomes easier for the leading polarities to cancel with the flux of the opposite polarity from the other hemisphere. Now consider a cycle that has become strong in which the BMRs appear in high latitudes. 
These high latitudes BMRs will produce less poloidal field and the next cycle will be weak. Hence, the indefinite growth of the magnetic field will be halted. \citet{Petrovay20} called this mechanism latitude quenching
and \citet{J20} argued that this mechanism could stabilize the growth of the magnetic field in the kinematic dynamo. 
\citet{Kar20} implemented this idea in a 3D \bl\ dynamo model by taking a simple latitude-dependent threshold for BMR eruption and showed that this latitudinal quenching can regulate the growth of the magnetic field when the dynamo is not too supercritical. 

\begin{itemize}
\item{\bf{Magnetic field-dependent inflows around BMRs}}
\end{itemize}
Surface observations show a converging flows around the BMRs \citep{Gizon01,Gonz08}.  These inflows 
cumulatively
generates mean flows around the activity belt whose strength depends on the amount of flux in the 
cycle \citep{Jiang10, CS12}. Due to these flows, the cross-equatorial cancellations of the BMRs are reduced and the effectivity of the \bl\ process is suppressed. In a strong cycle, this effect is stronger and thus lead to a stabilizing effect in the dynamo \citep{MC17, Nagy20}. 
  
In summary, tilt quenching, flux loss due to magnetic buoyancy, latitudinal quenching and  inflows are 
the potential candidates for the nonlinearity in the toroidal to the poloidal process of the \bl\ dynamo which can
potentially saturate the magnetic field.


\subsection{Time delay models}
\label{sec:timedelaymod}
As discussed in \Sec{sec:timedelay}, the time delay involved in various processes in the solar dynamo, operating concurrently with the nonlinearity can 
lead to irregular cycles.   
 \citet{Yo78} introduced a long delay of 29~years in the nonlinear dynamo model and found 
 occasional eras of suppressed activity. Although such a long delay is not expected in the solar dynamo, a finite delay 
 arises naturally in any dynamo model as long as the sources for the poloidal and toroidal fields are spatially segregated. 
 In the \bl\ dynamo models, the poloidal field after it is produced near the surface through the decay of tilted BMRs needs to be transported to the deeper CZ (through \mc, turbulent diffusion and pumping)
 where the toroidal field is generated through the ${\rm\Omega}$ effect. There is also a short time delay between the toroidal and poloidal field conversion as the toroidal flux tubes take finite time to rise to the surface to form BMR and a finite time is spend to decay and disperse the BMR.  All these two delays in the nonlinear dynamo model can produce a variety of modulations.

\subsubsection{Iterative map}
\cite{Dur00} assumed that there is a delay of one cycle between the poloidal field of cycle $n$ and the toroidal field of cycle $n+1$ and brilliantly reduced the \bl\ dynamo equations into a iterative map. He writes, 
\begin{equation}
   T_{n+1} = {\rm \Delta} {\rm \Omega} {\rm \Delta} t P_n = a P_n, ~~~~~~~~ n = 0, 1, 2, ...
   \label{eq:Durney1}
\end{equation}  
(Here $ {\rm \Delta} {\rm \Omega} $ is the shear in the CZ and  ${\rm \Delta} t$ is the time interval during which the poloidal field acts on the shear.)
Neglecting the time delay in the toroidal to poloidal fields conversion, we can write the following nonlinear relation: 
\begin{equation}
   P_{n+1} = f(T_{n+1}) T_{n+1}
   \label{eq:Durney2}
\end{equation}  
Here $f(T_{n+1})$ is a measure of the efficiency of the poloidal field generation from the toroidal field (\bl\ process) which depends on the toroidal field.  Substituting \Eq{eq:Durney1} into (\ref{eq:Durney2}) and normalizing the fields appropriately, we find
\begin{equation}
   p_{n+1} = a f(p_{n}) p_{n}
   \label{eq:Durney3}
\end{equation}  
(where normalizing factors are absorbed in $a$).
For different nonlinear functions ($f$) of the \bl\ process, different maps can be constructed. 
\citet{Dur00} chose it  $1 + \beta ( 1 - p_n)$ and thus the map became
\begin{equation}
   p_{n+1} =  p_{n} \left( 1 + \beta ( 1 - p_n) \right) ~~~~~~~~~~~~~ \beta > 0.
   \label{eq:mapDurney}
\end{equation}  
\citet{Cha01} chose it $\gamma (1 - p_n) p_n$ which produced a map 
\begin{equation}
   p_{n+1} = \gamma ~ p_{n} ^2 ( 1 - p_n) ~~~~~~~~~~~~~ \gamma > 0.
   \label{eq:mapCha01}
\end{equation}  
By capturing a lower cut off in the \bl\ process, \citet{CSZ05} produced another map
\begin{equation}
   p_{n+1} = \alpha f(p_n)  p_n ~~~~~~~~~~~~~ \alpha > 0,
   \label{eq:mapCha05}
\end{equation}
where
\begin{equation}
f(p_n) = \frac{1}{4} \left[ 1 + \mathrm{erf}\left(\frac{p_n - p_1}{w_1}\right)  \right] \left[ 1 - \mathrm{erf}\left(\frac{p_n - p_2}{w_2}\right)  \right]
\label{eq:low_qu}
\end{equation}
(Here $p_1 = 0.6$, $w_1 = 0.2$  $p_2 = 1.0$, and $w_1 = 0.8$.) 

\begin{figure}
\centering
\includegraphics[width=0.48\columnwidth]{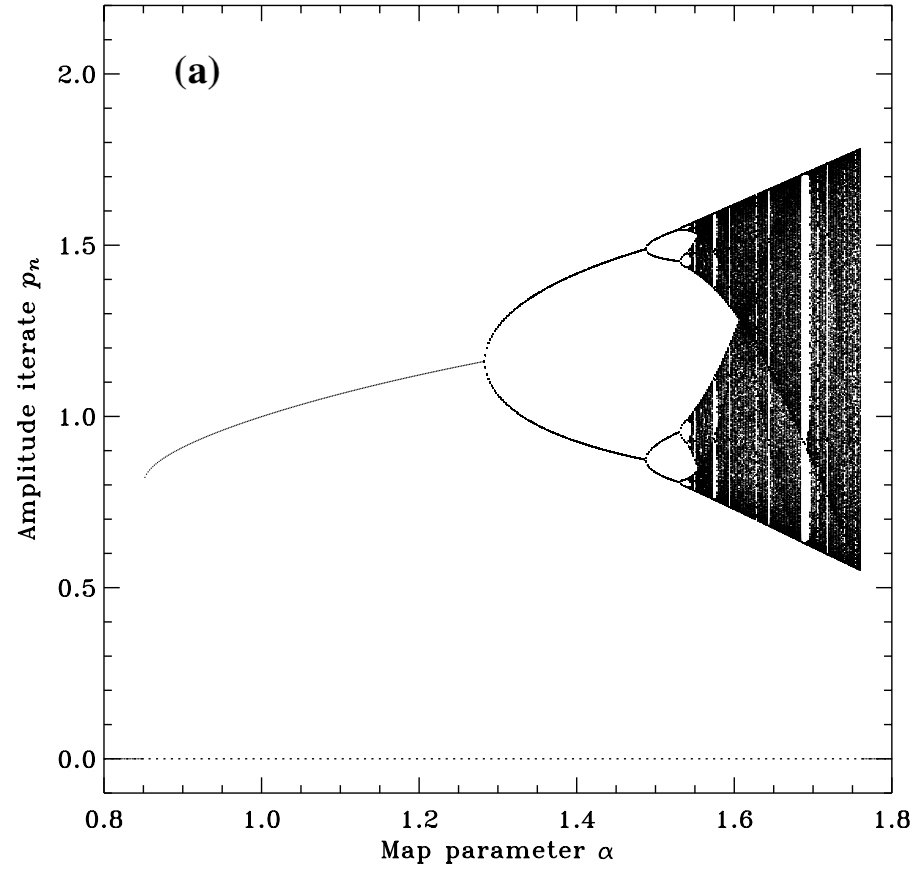}
\includegraphics[width=0.48\columnwidth]{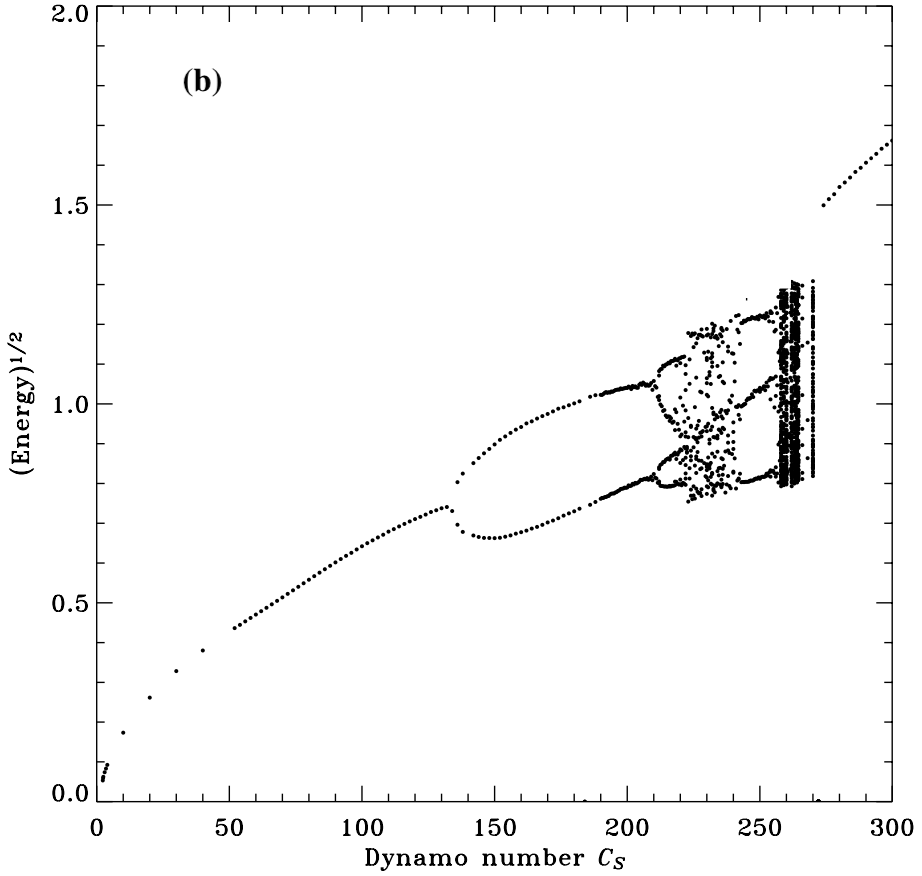}
\caption{Bifurcation diagrams. (a) Cycle amplitude (iterate $P_n$) vs the map parameter $\alpha$ from \Eq{eq:mapCha05}. (b) Same as left one but obtained from 2D numerical dynamo model of \citet{CSZ05} and shows the magnetic energy as function of the dynamo number $C_s$. Figures are reproduced from \citet{CSZ05} with permission from authors, copyright by the American Astronomical Society.}
\label{fig:map}
\end{figure}

\citet{Cha01, CSZ05} showed that as the map parameter increases, the transition from the fixed amplitude oscillation to the chaotic solution occurs through a sequence of period doubling; see left panel of \Fig{fig:map} for the map given by \Eq{eq:mapCha05}. \cite{Dur00} showed that in the parameter regime of the doubly periodic oscillations, the \go\ rule can be explained. 
Later \citet{Cha01} showed that this is indeed not necessary, stochastic perturbation outside this region can also produce \go\ rule as a consequence of the oscillatory nature of the convergence to the fixed point; even the map without showing limit cycle for $1 / (1 + B^2)$ type nonlinearity also show \go\ rule.
Intermittent and chaotic solutions are produced in all these maps, as long as the map parameter is above a certain value; also see Fig.\ 5 of \citet{Cha01} for an illustration.

\subsubsection{1D time delay dynamo}
Finite delays are also included in the 1D dynamo models in which the 
equations for the toroidal and poloidal fields are truncated by removing the spatial dependences in the following way.
\begin{equation}
\frac{\mathrm {d} B}{\mathrm{ d} t} = \frac{\omega}{L} A (t - T_0) - \frac{B}{\tau_d}
\end{equation}
\begin{equation}
\frac{\mathrm {d} A}{\mathrm{ d} t} = \alpha_0 f(B(t - T_1)) B(t-T_1)  - \frac{A}{\tau_d}.
\label{eq:delay2}
\end{equation}
Here, $T_0$  and $T_1$ represent the time delays required for the generation of toroidal and poloidal fields, respectively.  
 $\omega$ and $L$ are the contrast in
differential rotation and the length scale in the tachocline, $\tau_d$ is the diffusion time scale, $\alpha_0$ is the amplitude of the \bl\ source. $f(B(t - T_1))$ is the nonlinear function which represents the suppression of the \bl\ mechanism.  By considering $f(B(t - T_1))$ of the from of \Eq{eq:low_qu} (with lower quenching), \citet{wilsmith} found irregular cycles in a certain parameter regime (when the time delay is larger than the diffusion time). 
Later, including fluctuations in the $\alpha_0$ term, \citet{Ha14, kumar21b, tripathi21} obtained long-term modulations and grand minima like intermittent solutions in a range of parameters.

\subsubsection{2D time delay dynamo}
In models like the \bl\ type flux transport and the interface dynamos, in which the source regions for the fields are spatially segregated, the time delays are by default inbuilt into the equations. Thus all the \bl\ dynamo models discussed in this review are also time delay models. \citet{CSZ05} showed that the 2D \bl\ dynamo models also show the same type of behaviour as seen in the reduced map. \Fig{fig:map}(b), shows the bifurcation diagram of a 2D \bl\ dynamo model with a nonlinear quenching function of the form given by \Eq{eq:low_qu}. Again, with this type of nonlinearity, we observe that the solution goes to a chaotic regime through a sequence of period doubling with the increase of dynamo number. The model is capable to produce \go\ rule in a wide range of parameter regimes with stochastically forced $\alpha$ \citep{Cha07}; more in \Sec{sec:origin_OErule}. 
 

\begin{figure}
\centering
\includegraphics[width=1.0\columnwidth]{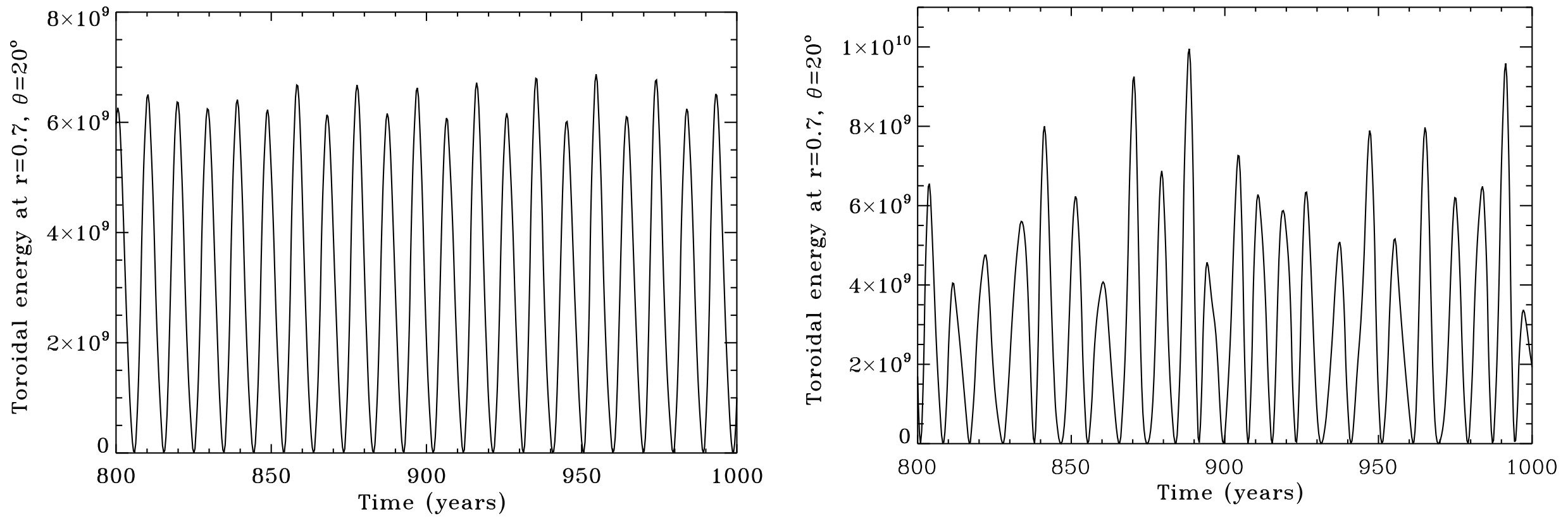}
\caption{Cycle modulations (as measured by the toroidal field at $r = 0.7R_{\odot}$ and $\theta = 20^\circ$) in the flux transport dynamo model of \citet{Jouve10} with magnetic field-dependent delay in the \bl\ source for the poloidal field generation process (\Eq{eq:delaypol}). The Left and right panels are for 
short (14 days on 1 kG fields) and long delay (14 days on 50~kG fields so that 10 kG fields will be delayed by almost a year in this case compared to a few hours in the previous case).
Image reproduced with permission from authors.}
\label{fig:jouve}
\end{figure}

\citet{Jouve10} went one step ahead of this and included the short time delay associated with the flux emergence from the deep-seated toroidal flux that is usually ignored in the dynamo models. They realized the fact that the buoyancy time delay depends on the magnetic field strength---strong flux tubes experience strong buoyancy and thus rise quickly compared to the weaker ones \citep{FFM94}. 
\citet{Jouve10} captured this delay in their flux transport dynamo model by replacing the poloidal source term: $\alpha B$ 
in \Eq{eq:pol} by 
\begin{equation}
 \frac{\alpha {B}(0.7R_\odot, \theta, t - \tau_{\rm B})} {1 + \left({ {B}(0.7R_\odot,\theta, t - \tau_{\rm B}) }/{B_0}\right)^2}, 
\label{eq:delaypol}
\end{equation}
where $\tau_{\rm B}$ is the delay time which they took to be equal to $\tau_0 / B(0.7R_\odot, \theta, t)^2$
 and $\tau_0$ is used to regulate the amount of delay. 
 \Fig{fig:jouve} shows the results from  two simulations
 with different amounts of delay as presented by \citet{Jouve10}. We evidently see that just the addition of this delay in the poloidal source produces a considerable amount of modulation in the cycle and the amount of modulation increases with the increase of
 delay \citep[also see][]{Fournier18}.
 
However, we note that in these models, the flux loss due to magnetic buoyancy is ignored (\Sec{sec:BLnonlinear}). 
The flux loss tries to keep the magnetic field around the equipartition value. In that case, the toroidal flux tube will not have much different field strength from one another and the delay times will not be very different. Hence, if the flux loss due to magnetic buoyancy is incorporated in these models then  
we do not expect much modulation in the cycle.  \citet{BKC22} included flux loss due to 
magnetic buoyancy in the dynamo model with local $\alpha$ prescription 
and they did not find noticeable modulation in the cycle due to field-dependent delay with respect to the case without delay.

\section{MHD simulations for long-term cycle variabilities}
\label{sec:MHDmodels}
In the MHD simulations one needs to solve the following continuity equation for the mass and the
energy equation in addition to \Eqs{eq:ind}{eq:mom}.
\begin{equation}  
  \label{eq:continuity}
 \frac{\partial \rho}{\partial t} +  \del \cdot( \rho \vec{v}) = 0,
\end{equation}
 \begin{equation}  
  \label{eq:heat}
 T \left[ \frac{\partial s}{\partial t} + (\vec{ v} \cdot \del ) s \right] = \frac{1}{\rho} \left[ \eta \mu_0 \vec{J}^2  -  \del \cdot (F_{\rm rad} + F_{\rm SGS}) - F_{\rm cool} \right] + 2 \nu \vec{S}^2,
\end{equation}
where $\rho$ is the density, $\nu$ is the kinematic viscosity, $F_{\rm rad}$ is the radiative diffusive flux 
and is given by $- K \del T$, $K$ being the heat conductivity, $F_{\rm SGS}$ represents the additional 
subgrid scale (SGS) diffusion which is used to keep the simulation numerically stable (usually taken as 
$-\chi_{\rm SGS} \rho T \del s^\prime$ with $\chi_{\rm SGS}$ being the SGS diffusion coefficient and $s^\prime$ is the fluctuations of entropy), and  $F_{\rm cool}$ is the radiative cooling near the surface. These equations 
are numerically solved using appropriate initial and boundary conditions to study the dynamo problem; see for example \citet{Kap20} for the detailed profiles of all the model parameters and the boundary conditions. 

\begin{figure}
\centering
\includegraphics[width=1.0\columnwidth]{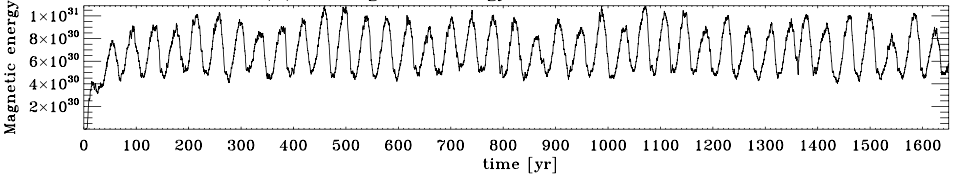}
\caption{Cycles from the global MHD convection simulations using EULAG-MHD code. Reproduced from \citet{PC14} with permission from authors.}
\label{fig:PC14}
\end{figure}

In the last one decade, global MHD convection simulations have reached to a somewhat realistic level (not in terms of the Reynolds numbers but in terms of the level of turbulence and the realistic value of Rossby number). They produced some
basic features of the large-scale flows and the magnetic field. We refer the readers
to Section~6 of \citet{Cha20} for a review on this subject.  An advantage of  these simulations is that all the nonlinear and stochastic effects are included by default in these simulations, in contrast to 
the mean-field models where these effects need to be included by hand. 
However, 
due to limited computation facilities, global convection simulations were rarely run for a longer time
so that a long-term cycle modulations can be studied. Furthermore, being extremely complicated in nature,
identifying the mechanisms of the long-term modulations are not trivial. 
\citet{FF14, Kar15, viv18, viv19} presented some simulations which were run for a somewhat longer duration.
Here we discuss three important results for the cycle modulations gleaned from three different numerical codes. The first one is from \citet{PC14} who presented cycles from a simulation run of 1650 years
in which 40 excellent cycles with average period of 40 years were seen (\Fig{fig:PC14}). 
This simulation `broadly' 
reproduces some observed features of the solar magnetic cycle including the regular polarity reversal and 
dipole-dominated large-scale field. Interestingly a good amount of modulation in the cycle amplitude is naturally produced in this simulations. There is also a pattern for the Gnevyshev-Ohl rule (watch the peaks after $t=800$~yr) and a hint for the Gleissberg modulation in these cycles.  However, in this 1650 years of simulations no grand minimum or maximum is seen.
With respect to the solar observations, there are some discrepancies as well which include an in-phase variation of the poloidal and toroidal components, magnetic activity confined to high latitudes, 
and the low degree of hemispheric coupling. 

\begin{figure}
\centering
\includegraphics[width=0.60\columnwidth]{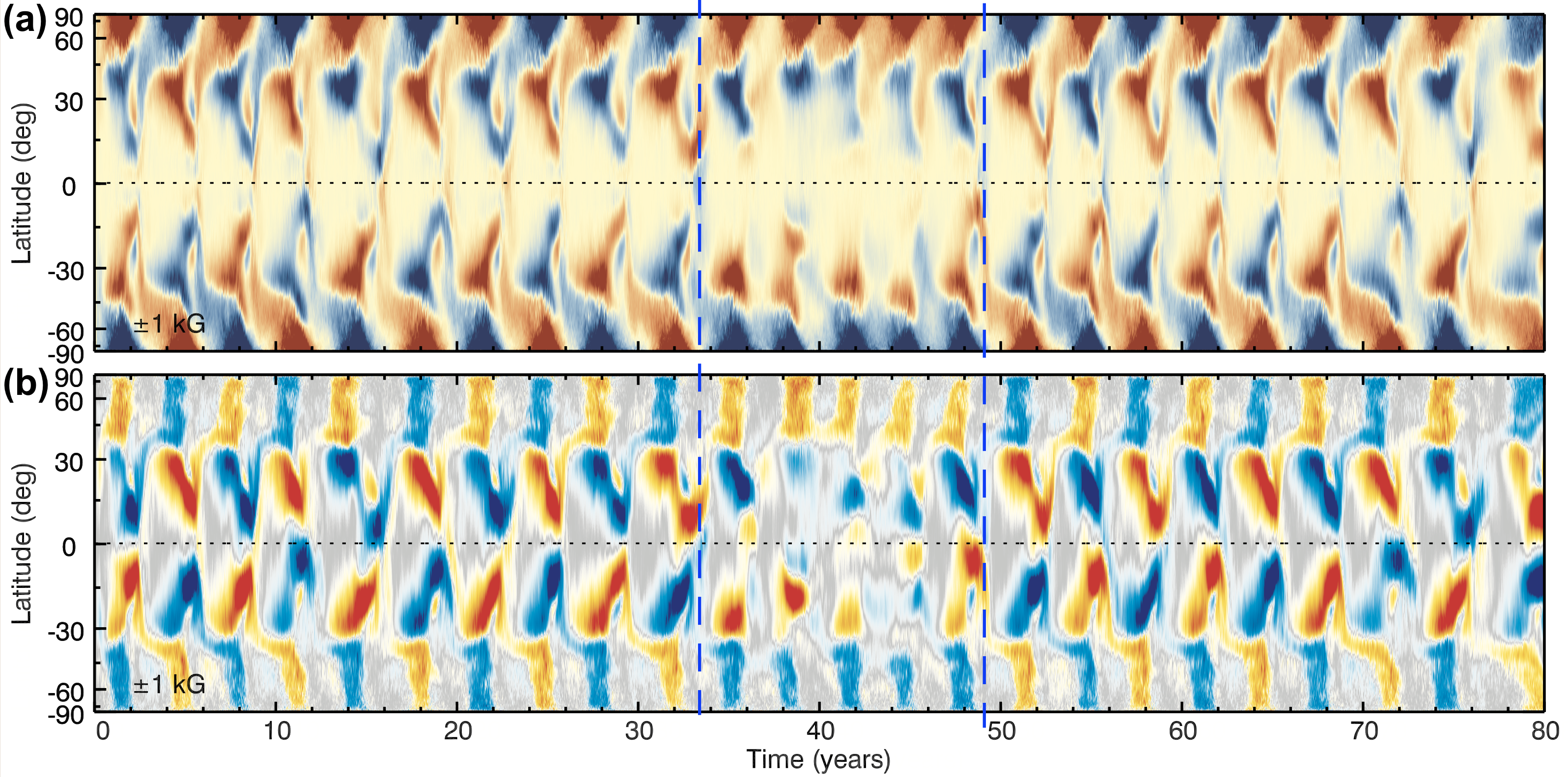}
\includegraphics[width=0.39\columnwidth]{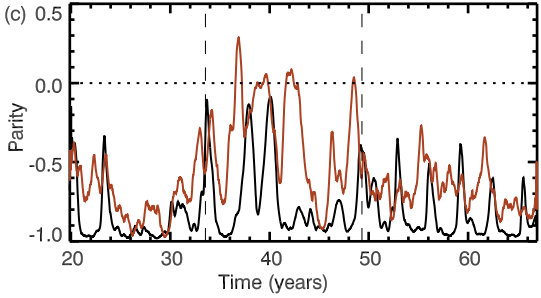}
\caption{Left top: Time-latitude distributions of the longitude average $B_r$ 
and $B_\phi$ at $0.92R_\odot$ from the ASH simulation code of \citet{ABMT15}. 
The grand minimum identified in this simulation is marked by vertical lines. Right panel shows the
parity of the magnetic field computed at $0.75R_\odot$ (orange curve) and $0.95R_\odot$ (black).
Figure reproduced with permission from authors, copyright by AAS.}
\label{fig:Aug}
\end{figure}

The second important result came from \citet{ABMT15} utilizing the 3D MHD ASH code 
which is shown in \Fig{fig:Aug}. 
This is from a model of one solar mass rotating at three times the Sun. Again this simulation produces many features that are consistent with observation and the important one is the equatorward migration of the toroidal field belt 
at low latitudes which is caused by the nonlinear modulation of the differential rotation.  
The interesting feature in their simulation is that a clear grand minimum
was identified with a considerably reduced magnetic field for about five cycles. 
However during this minimum, the large-scale field failed to reverse, although its amplitude 
oscillated cyclically. The grand minimum in this simulation is possibly caused by the interplay between the
symmetric and anti-symmetric dynamo families. During the regular cycle, the anti-symmetric dynamo 
family is greater than the symmetric family but during grand minimum phase, the symmetric family dominates over the anti-symmetric one; see the right panel of \Fig{fig:Aug} for the increased parity of the radial field at two different depths. The mechanism for the generation of the grand minimum in this simulation is similar to the 
one proposed by \citet{Tob97,MB00} based on the nonlinear mean-field dynamo model. 

\begin{figure}
\centering
\includegraphics[width=1.00\columnwidth]{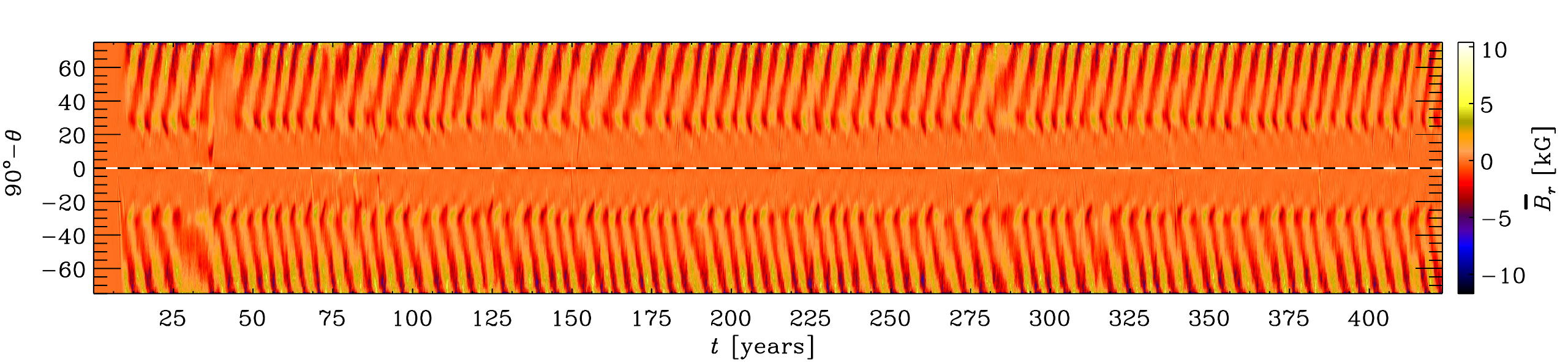}
\caption{Temporal variation of the mean $B_r$ average over the longitudes on the surface from global convection simulation using Pencil Code \citep{Kap16}. The color scale is saturated at the half of the extrema i.e., at [$-11.7$, 10.4]~kG. Note the disturbed magnetic activity during $t =$ 20--45 years.
}
\label{fig:kapyla}
\end{figure}

The final one is from \citet{Kap16} utilizing the Pencil Code. They performed this simulation for a model
of one solar mass rotating at five times the Sun. This simulation also produced some solar-like features, including
regular polarity reversals and an equatorward migration of the toroidal field at low latitudes (due to a nonlinear dynamo 
wave). In addition to the dominant global magnetic cycle of an average period of 4.9 years, 
there are two other prominent cycles, one having a higher frequency mode near the surface 
and at low latitudes with poleward migration, 
and the other one having low frequency residing at the BCZ. 
Their simulation also finds an episode of reduced magnetic field for about three cycles, happening asynchronously in hemispheres ($t =$ 20--45 years). Interestingly, the magnetic field in the deeper CZ is stronger during this period and thus the global magnetic field during this period 
is larger than during the normal phase.
 The dynamics of the magnetic fields in this simulation are extremely complex and 
 the grand episode of the reduced activity is caused through the interplay of various dynamo modes.

Although the global convection simulations produce some long-term modulations 
in the cycles and the grand minimum-like reduced activity which are consistent with solar observations, 
there are few caveats that we should keep in mind.  
The cycle modulations observed in the above convection simulations are many ways 
far from the actual Sun. The large-scale flows produced in these simulations are also
quite far from the real Sun. Importantly, the power in the sub-surface convective flow at large scales
are much stronger than that is obtained from the measurements, so-called the convective conundrum  (\citet{Hanasoge12, Lord14}; also see e.g., \citet{HRY15, KMB18b} for the studies that attempt to resolve this.) 
Furthermore, we do not have any information about whether these 
long-term modulations found in the convection simulations are robust in the model parameter regimes. Finally, the simulations do not produce BMRs which are important component of the solar cycle and are responsible for the generation of the poloidal field in the Sun, at least the field that is observed on the surface. 
Therefore, future work is needed to make the global convection simulations
more realistic so that they can be utilized to study the long-term cycle modulation.

Ideal MHD simulations in the local box are also performed to study the large-scale dynamos. 
With helical forcing and imposed shear, \citet{KKB15} performed HMD simulations in the local Cartesian geometry and found modulations in the large-scale magnetic cycle. The interesting 
fact about their study was that they found grand minima like intermittent activity only when the 
dynamo operates in the subcritical and critical regime but not in the supercritical regime. This
 independent study thus supports the idea that the variability and the grand minima are less 
probable in the supercritical dynamo; see \Sec{sec:supercriticality} for details.
Local simulations are also useful to study the cycle modulations in presence of the small-scale dynamo.
The small-scale magnetic field generated from the small-scale dynamo affects the flows and thus 
the global dynamo \citep{KB16}.
Global dynamos are usually performed at low Reynolds numbers and thus small-scale dynamo is not excited, 
except a few \citep[e.g.,][]{Nel13, Kap17, HK21}, however, they are not ran for many cycles.

\section{Some open questions and current trends}
\label{sec:questions}
\subsection{Do grand minima represent different states of the solar dynamo?}
\begin{figure}
\centering
\includegraphics[width=.5\columnwidth]{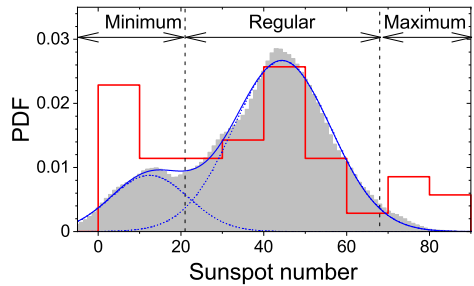}
\caption{Probability distribution function of the reconstructed sunspot number (filled grey curve) and the observed group sunspot number (red curve). Blue is the bi-Gaussian curve. Figure is modified after \citet{Uea14}; also see 
\citet{Wu18} for the distribution from longer data.}
\label{fig:distinctmode}
\end{figure}

By analyzing the solar activity data for the past, \citet{Uea14} showed that the distribution of 
the solar activity is bi-modal. Of this, the dominant mode corresponds to the regular activity phase 
and the reduced-activity mode which corresponds to the grand minima 
is distinct from the dominant regular one. As seen from \Fig{fig:distinctmode}, the distribution is clearly bimodal. 

In terms of the dynamo theory, of course, the physics during grand minima is not quite the same as that during the regular cycle. For example, during the grand minima phases, the magnetic field 
falls to a low value and then the Sun takes some time to grow its magnetic field to the normal level. During these grand minima phases, the generation of the poloidal field is low because the \bl\ process which is the dominant source for the poloidal field in the Sun becomes less efficient due to fewer BMRs (\Sec{doesBLoperate}). The strength and morphology of the flow can also be different during this phase. 
Therefore, the \bl\ dynamo models coupled with weak $\alpha$ effect and/or the dynamo models coupled with the dynamic $\alpha$ effect produced by the instability of flux tube (discussed in \Sec{sec:mod_coupledalpha}) can naturally explain the bimodal distribution.
The dominant source of the poloidal field (\bl\ process or dynamic $\alpha$ effect) maintains the normal mode of the solar activity while the grand minima phase is maintained by the weak $\alpha$ effect. Time delay models or the iterative map with low-amplitude additive noise and fluctuating map parameter/dynamo number with specific nonlinearity can also naturally produce the bimodal distribution of solar activity; see Figure 1 of \citet{Cha01} and \citet{tripathi21}. Also, see \citet{Pet07} for a possible explanation of a bimodal solar dynamo based on an interface dynamo model coupled with a fast tachocline model. \citet{KKB15} using 3D simulations of turbulent dynamo suggested that the bimodal distribution of solar activity can be produced in the subcritical dynamo.

\subsection{Do grand maxima require different mechanisms for their origin?}
In most of the previous studies, the origin of the grand maxima is ignored.
However, the mechanism for its generation is subtle because the dynamo is more nonlinear
during the grand maxima phase. When the Sun tries to produce a strong magnetic field, the nonlinearity tries to quench its generation process (efficiencies of both the poloidal and toroidal field generations are reduced with magnetic field). 
Incidentally, the Sun spent less time in the grand maxima phase than in the grand minima phase  \citep{USK07,sol04}. 
Thus, producing extended grand maxima using the stochastic fluctuations in the dynamo parameters is less obvious.
However, stochastic fluctuations still can produce a very strong magnetic cycles and grand maxima if they occur in a certain phase of the cycle. \citet{KO16} showed that at the beginning of a cycle (when the field at the poles is strongest) if the generation of the poloidal field is reversed (say due to the emergence of some wrongly tilted BMRs), 
then it will produce the same polarity field as it was there in the pole. Consequently, instead of reversing the old polarity polar field, it will amplify.  This strong polar field will make the current cycle very strong.  By introducing stochastic fluctuations in a 2D flux transport dynamo model, \citet{KO16} showed that this mechanism can occasionally produce a much stronger cycle which corresponds to the grand maxima phase. However, in this mechanism, not more than one strong cycle at a time is produced, while in the solar grand maximum, at least two consecutive cycles are strong \citep{USK07}.   Also, this study is based on a kinematic model, 
in which the nonlinear feedback of the magnetic field on the flow is ignored. 

Another way of generating the grand maxima is through the combined effect of multiple poloidal field generation processes.
\citet{olc19} find that when the deep-seated $\alpha$ effect is coupled with the surface \bl\ process in a dynamo model,  these
two processes more or less contribute equally to the generation of the poloidal field through a sort of constructive interference. This could be the mechanism of grand maxima in their dynamo model. However again this model 
is kinematic and the magnetic feedback is not taken care of.
 
 \subsection{What is the origin of Gnevyshev-Ohl/Even-Odd rule?}
 \label{sec:origin_OErule}
One plausible explanation for the Gnevyshev-Ohl rule is the fossil field hypothesis. 
 A steady large-scale magnetic field of fossil origin \citep{Boruta96} can interfere with the oscillating magnetic field from the CZ. In one cycle the oscillating magnetic field appears in the same polarity as that of the fossil field and it 
 makes the cycle strong. In the next cycle, the oscillating magnetic field becomes of opposite polarity and 
 thus the cycle becomes weak.  To explain the Even-Odd rule using this idea, the fossil field has to be
 comparable to the dynamo-generated oscillating magnetic field at the BCZ which is of the order of 10~kG.
 However, the present observations do not confirm this strong fossil field.
Furthermore, if the \go\ rule is caused by the fossil field,
then there should be infinite memory in it, in a sense that once this rule is established, odd (even) cycles will always be stronger than the previous even (odd) cycles, even if there are some violations due to other effects.  
However, studies show that there was a possible reversal in the even-odd pattern 
during 1745--1850 \citep{MUK01, Tlat13, ZP15}.

 There is another possible explanation for the Gnevyshev-Ohl rule which was proposed by \citet{Dur00} using 
the nonlinear period-doubling effect. He suggested that the solar dynamo is operating in the region of period doubling beyond the bifurcation point and in this region the alternating amplitude modulation is unavoidable.  
Later \citet{Cha01} showed that this is indeed not essential; stochastic forcing in the dynamo can lead to 
Even-Odd effect even outside this parameter range of period doubling. 
The time delays (\Sec{sec:timedelaymod}) involved in the solar dynamo including simple amplitude limiting nonlinearity can produce the same period-doubling and Gnevyshev-Ohl rule as seen in the complex nonlinear system \citep{Cha07}. The dynamo model and map with various types of nonlinearity show \go\ rule under fluctuations in the poloidal source ($\alpha$). 
However, this \go\ rule becomes evident only when the nonlinearity becomes important (large map parameter or dynamo number) and it seems to be little restricted to a narrow range of the diffusivity in the BCZ 
\citep[Sec 4.5 of][]{CSZ05}.
Thus, there is still room to explore the robustness of this feature.


 \subsection{What are the causes of Gleissberg and Suess/de Vries cycles?}
Gleissberg and Suess/de Vries cycles are not strictly cycles, rather they are modulation over the dominant 11-year period
and they are detected in the cosmogenic data having ranges of periods from 90 to 100 years and from 205 to 210 years, respectively.
If these modulations are the true nature of the solar cycle, then they are probably coming from the nonlinear interaction between the 
magnetic field and the flows in the lower part of CZ. 
In a axisymmetric $\alpha$$\Omega$ dynamo model coupled with the angular momentum equation and ignoring \mc, \citet{Pip99} showed that the Gleissberg cycle is a results from the 
magnetic feedback on the angular momentum fluxes which maintains the 
differential rotation in the CZ. In this model, the period of the Gleissberg cycle is determined by the time associated with the re-establishment of the differential rotation after the magnetic perturbations of the angular momentum transport.
If this is the mechanism of the Gleissberg cycle in the sun, then the observed differential rotation should show a variation in the Gleissberg timescale. 
However, the available observations does not provide a conclusive evidence on it \citep{How78}.
\citet{PC14} found a hint of Gleissberg modulation in their computed proxy sunspot number but not in the radial     
field data and they require data for a longer duration to confirm its existence.
\citet{CS19} showed that Gleissberg and Suess/de Vries cycles are consistent with realization noise and the noisy normal form model can reproduce these modulations.

 \section{Summary and discussion}
 \label{sec:conclusion}
  Besides the 11-year (a)periodic variation of the amplitude, the most prominent variation of the solar cycle is the long-term modulation which has been unambiguously identified in the direct and indirect (cosmogenic data) observations of the solar activity. 
 Examples of the long-term modulation include the Gnevyshev-Ohl/Even-Odd rule, \gm, grand maxima, Gleissberg cycle and Suess cycles. In this review, we have presented comprehensive discussions on the origins and models 
 of long-term variations. To do so, we have broadly identified the following three major causes for the cycle modulations: (i) magnetic feedback on the flow, (ii) stochastic forcing, and (iii) time delays in various dynamo processes. Problems in the nonlinear mean-field models are that not all possible nonlinearities are always included in the model and the resulting mean flows and magnetic fields are not compared carefully with observations. 
Global MHD convection simulations are very useful in this respect because all nonlinearities are captured by default. They have begun to produce some correct results of the large-scale field and the flow, however, there are big discrepancies as well. 
 Until global convection simulations reach somewhat realistic parameter regimes and resolve the 
 major issues (convective conundrum, observed large-scale flows, nonappearance of BMRs), we need to rely on the mean-field models only. 
 
Stochastic fluctuations in the mean-field models are enough to explain many features of the long-term variabilities. The \bl\ dynamo models are promising models for the solar cycle, in terms of their success in reproducing the long-term modulations in the solar cycle. Stochastic fluctuations in these models are due to randomness in the BMR properties (primarily due to scatter around Joy's law, BMR emergence rates and emergence latitudes). \bl\ models are also nonlinear because at least the toroidal to poloidal field generation step includes several essential nonlinearities (tilt quenching, latitude quenching, toroidal flux loss due to magnetic buoyancy). 
While these nonlinearities have the tendency of stabilizing the magnetic field, they can lead to fluctuating cycles including \go\ rule and grand minima at highly supercritical regimes (large dynamo numbers) due to the inherent time delay in the dynamo models with spatially segregated source regions.   
Observations indicate that the solar dynamo is possibly operating in a weakly nonlinear regime (slightly above the dynamo transition) and thus cycle modulations are caused by stochastic effects.  

One way to pick up the correct model out of all the possible models for long-term modulation 
is to carefully compare the model results with the observations. 
Observational results include the followings (but not limited to).
(i) There is a strong hemispheric asymmetry during the second half of the Maunder Minimum \citep{RN93}.
(ii) However, the asymmetry during the normal cycle is less, appears randomly, and is smoothed out in a few
cycles \citep[the memory of the asymmetry does not remain for multiple cycles;][]{GC09,McInt13,das22}.
(iii) The differential rotation in the whole CZ of the Sun is well-measured for about last four decades and it has only a tiny variation \citep[surface differential is measured even for about 200 years and also shows little variation;][]{GH84,Jha21}.
 
Grand minima are possibly triggered by the stochastic fluctuations in the dynamo parameters or/and the nonlinear interaction of the Lorentz force. The recovery to the normal phase from grand minima is trivial in any model which includes the $\alpha$ effect because it can operate in the weak-field regime. Recovery through the \bl\ process is also possible because recent analyses revealed spots during Maunder minimum and the \bl\ process can produce a poloidal field with a few BMRs or smaller BMRs having nonzero tilts (including ephemeral regions).
Grand maxima are probably more special events and less frequent than grand minima. Again nonlinear modulation of the flow via Lorentz force and stochastic fluctuations can trigger these events. A dual source of poloidal field generation occurring constructively or reversed generation of the poloidal field due to stochastic fluctuations can produce prominent grand maxima.  Occurrences of grand minima and maxima
in the dynamo models can be described by stochastic processes and these are consistent with the observations. The waiting time distributions of the grand minima and maxima in the dynamo models are also described by the memoryless stochastic processes, which however disagree with the available observations.

{\bf{Acknowledgments}}
Author shows gratitude to Akash Biswas, Bibhuti Kumar Jha, Leonid Kitchatinov, Paul Charbonneau, Pawan Kumar, and 
two (anonymous) brilliant referees for carefully reviewing and pointing out many mistakes in the earlier version of the draft. Author also thanks Pawan Kumar and Anu B Sreedevi for their help in preparing many figures.
Ilya Usoskin, Leonid Kitchatinov, Jie Jiang, Alexandre Lemerle, Maarit K\"apyl\"a, Dibyendu Nandi, and Sacha Brun have kindly provided data and/or figures for this article.
Author acknowledges financial support provided by ISRO/RESPOND (project No. ISRO/RES/2/430/19-20) and Ramanujan Fellowship (project no SB/S2/RJN-017/2018)
and the computational resources of the PARAM Shivay Facility under the National Supercomputing Mission, the Government of India, at the Indian Institute of Technology Varanasi.
SOHO is a project of international cooperation between ESA and NASA.
Courtesy of NASA/SDO and the HMI science teams.

\bibliography{paper}

\end{document}